\documentclass[aps,twocolumn,showpacs,preprintnumbers,amsmath,amssymb,superscriptaddress,prl,floatfix,10pt]{revtex4-2}

\usepackage{graphicx}
\usepackage{dcolumn}
\usepackage{bm}
\usepackage{subfigure}
\usepackage{epsfig}
\usepackage{epstopdf}
\usepackage{xcolor}
\usepackage{multirow}
\usepackage{mathtools}
\usepackage{nicefrac}
\usepackage{soul}
\usepackage{orcidlink}
\usepackage{hyperref}
\hypersetup{colorlinks=true,citecolor=blue,linkcolor=blue,urlcolor=blue}
\soulregister\eqref7
\soulregister\mbox7

\allowdisplaybreaks

\begin{document}

\title{Probing Fractional Quantum Hall States in Weakly Interacting Fermi Gases}

\author{Viktor Bekassy\,\orcidlink{0000-0002-7252-9144}}
\email{bekassy@chalmers.se}
\affiliation{Department of Microtechnology and Nanoscience (MC2), Chalmers University of Technology, 41296 Gothenburg, Sweden}

\author{Mikael Fogelstr\"om\,\orcidlink{0000-0003-3441-2559}}
\email{mikael.fogelstrom@chalmers.se}
\affiliation{Nordita, Stockholm University and KTH Royal Institute of Technology, 10691 Stockholm, Sweden}
\affiliation{Department of Microtechnology and Nanoscience (MC2), Chalmers University of Technology, 41296 Gothenburg, Sweden}

\author{Johannes Hofmann\,\orcidlink{0000-0002-0667-2452}}
\email{johannes.hofmann@physics.gu.se}
\affiliation{Department of Physics, Gothenburg University, 41296 Gothenburg, Sweden}
\affiliation{Nordita, Stockholm University and KTH Royal Institute of Technology, 10691 Stockholm, Sweden}

\date{\today}

\begin{abstract}
Quantum gases are used to simulate the physics of the lowest Landau level with neutral atoms, which in the simplest setup is achieved by rotating the gas at the confining harmonic trap frequency, a requirement that is difficult to meet in practice. 
However, it is known that lowest Landau level states appear among the excited states of weakly interacting gases, even in a nonrotating trap, by working at fixed finite angular momentum. Thus, even in the absence of rotations or for moderate rotation frequencies, fractional quantum Hall (FQH) states can be accessed as excited states at finite angular momentum. Motivated by this, and by recent experimentally demonstrated excitation protocols for two particles, we show that such states can be selectively addressed at larger particle numbers, and we propose imaging techniques to resolve their distinct microscopic structure. 
In particular, we suggest that recently developed excitation and imaging techniques for rotating few-fermion systems allow for a detailed experimental investigation of FQH wave functions and of the crossover to large particle numbers. We illustrate the evolution of the excitation spectrum with rotations at weak interactions for \mbox{$N=6$} spin-balanced fermions and show how sampling of the few-body wave function allows one to discriminate between total guiding-center, center-of-mass vortex, and hyperradial excitations, thus providing detailed and versatile probes of the microscopic structure of FQH states.
\end{abstract}

\maketitle

Since the first realization of Bose-Einstein condensation~\cite{Anderson95,davis95}, rotating ultracold atomic gases have been used to emulate magnetic fields~\cite{cooper08,bloch08}, beginning with the observation of a single quantized vortex~\cite{madison00} and Abrikosov lattices~\cite{Shaeer01}. Here, the Coriolis force on neutral atoms in the rotating frame acts in the same way as the Lorentz force on a charged particle, with an effective magnetic field that is proportional to the rotation frequency~$\Omega$~\cite{bloch08}. Seen classically, the atoms in a rotating trap perform a Foucault pendulum motion with a slow guiding-center oscillation with frequency \mbox{$\omega - \Omega$} on which a fast counterrotating cyclotron motion with frequency $\omega + \Omega$ is superimposed, where $\omega$ is the trapping frequency of the harmonic oscillator potential~\cite{crepel23}. However, reaching the limit of fast rotations~\mbox{$\Omega \to \omega^{-}$}, where the guiding-center frequency vanishes and quantum states form degenerate Landau levels that are separated by an effective cyclotron frequency~\mbox{$\omega_c = 2\omega$}, has only recently been possible in dynamical experiments in which the expansion dynamics of a rotating unstable trap ``geometrically squeeze'' the gas into the lowest Landau level (LLL)~\cite{fletcher21,mukherjee22,crepel23}. In equilibrium, rapidly rotating gases suffer from strong particle losses~\cite{viefers08,dalibard11,dalibard15}, and other challenges exist for alternative schemes to create artificial magnetic fields~\cite{aidelsburger11,struck12,Lin09,nascimbene25}. Yet, probing this limit is desirable since it would make it possible to test fractional quantum Hall (FQH) states in a very controlled setting, including few-particle systems~\cite{gemelke10}. 

In this context, several recent experiments have succeeded in creating {\it two}-particle FQH Laughlin states with photon pairs~\cite{clark20}, two bosons in a Floquet-engineered optical lattice~\cite{leonard23}, as well as two interacting fermions in an optical tweezer~\cite{lunt24}, the latter creating a generalization of the Laughlin state for two-component systems, a Halperin state~\cite{halperin83}. The key question is whether such approaches scale to larger particle numbers, in which case emergent topological properties could be probed with single-particle resolution. For example, a three-boson Floquet-engineered FQH state has recently been realized~\cite{kwan26}, and proposals exist for optimal control strategies to increase the particle numbers~\cite{wu25}. For rotating gases, existing theoretical proposals assume an adiabatic transfer from the nonrotating $N$-particle ground state to states at fast rotations~\mbox{$\Omega \simeq \omega$} induced by a gradual increase of the rotation frequency, which is followed by a transfer between states within the LLL~\cite{baur08,palm20,popp04,andrade21}; yet keeping particles trapped in practice is difficult due to the centrifugal force~\cite{bloch08}.

In this Letter, we show that the specific excitation protocol realized for two particles in Ref.~\cite{lunt24} can be used to reach targeted many-body LLL states for larger particle numbers, and we propose that single-particle imaging can resolve the resulting states' distinct microscopic structure.
This relies on the fact that LLL states appear among the excited states of weakly interacting gases, even in a nonrotating trap, by working at fixed finite angular momentum, which was established in Refs.~\cite{wilkin98,cooper08}. 
Weak here refers to a parameter regime in which the energy splitting due to the contact interactions (or generally any interaction that conserves angular momentum) is small compared to the harmonic oscillator energy $\hbar \omega$, a perturbative regime readily accessible in experiments~\cite{bayha20,lunt24}. Crucially, the criterion for the applicability of perturbation theory is independent of the rotation frequency $\Omega$, and, seen in the limit \mbox{$\Omega \to \omega^{-}$}, becomes the well-known criterion to avoid transitions between Landau levels, which are separated by the effective cyclotron energy \mbox{$2\hbar \omega$}~\cite{yannouleas20,hofmann23}. For Bose gases, this implies that LLL states form the ground-state manifold for perturbative interactions~\cite{wilkin98}, which also holds for Fermi gases at sufficiently large angular momentum~\cite{cooper08}. Hence, Fermi gases that are equilibrated at rapid rotations can realize LLL ground states, including FQH states~\cite{Regnault04}. The correspondence follows from first-order degenerate perturbation theory together with angular momentum conservation: The interaction lifts the degeneracy of states sharing noninteracting energy and angular momentum, with wave functions that are independent of the interaction strength~\cite{sakurai94}. 
Because of angular momentum conservation, rotating the gas only rearranges the energy spectrum while leaving the states themselves unchanged. 
Perturbation theory in the interaction parameter for certain excited states is thus equivalent to exact diagonalization within the LLL~\cite{cooper08}.
Here, we show that this correspondence has a direct impact on current experimental proposals and allows probing of ground and excited LLL states in existing experimental setups beyond two particles.

Our discussion is particularly motivated by the recent experiment by \textcite{lunt24}, which creates a two-fermion Halperin state by imparting an angular momentum~\mbox{$m \hbar = 2\hbar$} to the two-particle ground state in an effective two-dimensional harmonic trap. In the experiment, this is achieved by Laguerre-Gauss laser beams, which create a single-particle perturbation of the form
\begin{align} \label{eq:excitation}
\delta H_m \sim z^m e^{-i\Omega_{p} t}  +\text{c.c.}
\end{align}
Here,~\mbox{$z=x+iy$} is a complex particle coordinate, and the pulse frequency~\mbox{$\Omega_{p} \approx 2 \omega$} is chosen to create an excitation to an energy level with total angular momentum~\mbox{$M=2$}~\cite{lunt24b}. Without interactions, this excited level is threefold degenerate since the angular momentum can be imparted to the center of mass (c.o.m.), the relative degree of freedom, or both (the latter is a triplet state that is not excited by $\delta H_m$, which conserves total spin). Formally, the excited relative state is a Halperin state, $\Psi_{(m_1,m_2,m_3)}$ with \mbox{$(m_1,m_2,m_3) = (0,0,2)$}~\footnote{The Halperin wave function is given by \mbox{$\Psi_{(m_1,m_2,m_3)}(z_{1\uparrow}, \ldots, z_{1\downarrow},\ldots) =\prod_{i<j=1}^{N_{\uparrow}}  (z_{i\uparrow} - z_{j\uparrow})^{m_1}$} $ \prod_{i<j=1}^{N_{\downarrow}} (z_{i\downarrow} -z_{j\downarrow})^{m_2}\prod_{i,j=1}^{N_{\uparrow}, N_{\downarrow}}  (z_{i\uparrow} - z_{j\downarrow})^{m_3}$, and the two-particle state discussed in the text corresponds to \mbox{$N_{\uparrow}=N_{\downarrow}=1$} and \mbox{$(m_1,m_2,m_3) = (0,0,2)$}.}. The repulsive contact interaction separates these excited states, which allows a complete and selective population transfer to the Halperin state 
by adjusting $\Omega_p$. A main implication of our discussion is that schemes such as the one described for two particles should readily extend to larger ensembles and general states, even when the initial ground state is not within the LLL. This offers a complementary way of realizing FQH states. Combined with the ability to experimentally sample few-particle wave functions~\cite{bergschneider18,bayha20,holten21a,holten21b}, this puts an experimental study of few-body FQH physics and perhaps even the crossover to the many-particle limit within reach.

This Letter is structured as follows. We begin by describing the single-particle spectrum in a trap, how levels rearrange with rotations, and the relation to degenerate perturbation theory. Next, we illustrate the rearrangement of the many-body excitation spectrum from slow to fast rotations, where LLL states form the lowest-lying excitations. By way of example, we discuss the spectrum of a spin-balanced two-component Fermi gas with~\mbox{$N=6$} particles and consider the minimal-angular-momentum LLL states excitable from the nonrotating ground state using the scheme [Eq.~\eqref{eq:excitation}]. Finally, we propose how spin-resolved single-particle imaging techniques used in current experiments~\cite{bergschneider18,holten21a,holten21b,lunt24,lunt24b} could provide a novel way to observe and distinguish these different final states by sampling the many-body wave function.

\begin{figure}[t!]
    {
    {\includegraphics{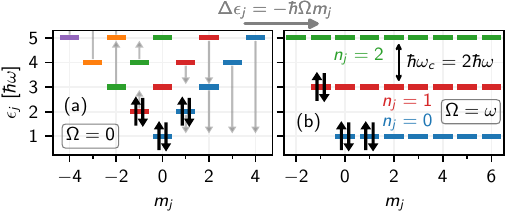}}}
    \caption{Single-particle energy spectrum in a 2D harmonic trap as a function of angular momentum~$m_j$ (a)~at rest and (b)~in the frame of rapid rotation, where the color indicates the number of cyclotron excitations. In a rotating frame, energies are shifted proportionally to $m_j$ (gray arrows) but the wave functions remain unchanged. For rapid rotations, states form degenerate Landau levels labeled by $n_j$. Noninteracting few-body states are obtained by successively populating single-particle states, illustrated here for the ground state of~\mbox{$N=6$} fermions in a stationary trap, which at rapid rotation is an excited state due to the population of a cyclotron excitation.
    }    
    \label{fig:1}
\end{figure}

\begin{figure*}
    { 
{\includegraphics{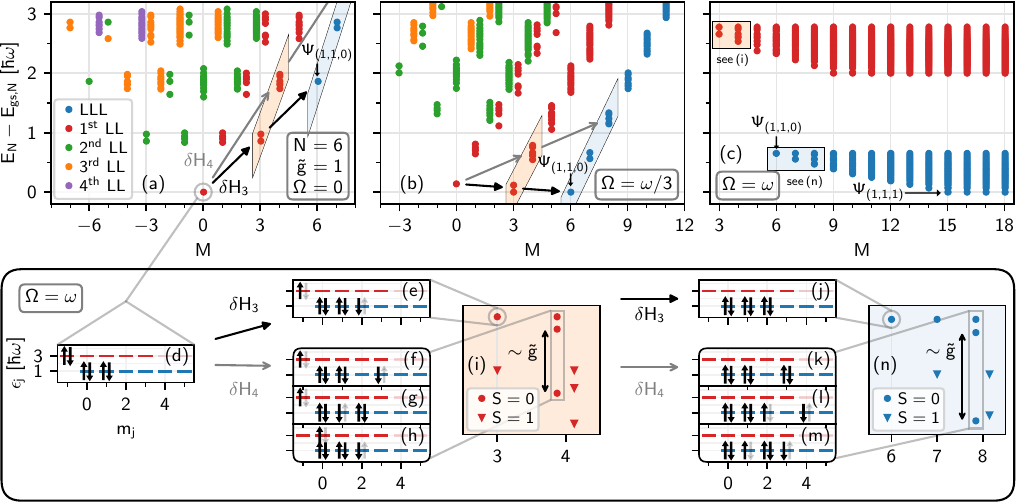}}}
    \caption{
    Upper row [panels (a)--(c)]: excitation spectra of~\mbox{$N=6$} fermions as a function of total angular momentum $M$ for rotation frequencies~\mbox{$\Omega=0,\omega/3$}, and $\omega$ for a repulsive perturbative interaction~\mbox{$\tilde{g}=1$}. Colors indicate the highest populated single-particle Landau level in the $N$-particle state with the same color coding as in Fig.~\ref{fig:1}, and overlapping states of different color are offset horizontally. The key point, discussed in detail in the main text, is that while LLL states form the lowest-lying excited states as \mbox{$\Omega \to \omega$} [blue points in panel~(c)], they are always present, even in a nonrotating trap, as excited states. Black and gray arrows indicate a possible two-step excitation to selected LLL states (blue shaded) that can be reached from the nonrotating ground state through the perturbation~\mbox{$\delta H_m$} [Eq.~\eqref{eq:excitation}] via states highlighted in orange. The lower row shows this excitation path in more detail, visualizing the population of states for \mbox{$\Omega = \omega$}.
    (d) Nonrotating ground state (with total spin \mbox{$S=0$}) as in Fig.~\ref{fig:1}. (e),(f) States excited by \mbox{$\delta H_{3}$} and \mbox{$\delta H_{4}$}, which transfer one particle from the first LL to the LLL by imparting angular momentum \mbox{$m=3$} and \mbox{$m=4$}, respectively. The noninteracting state~(f) is degenerate with the states (g) and (h) [unpaired spins form a singlet or triplet with their spin-flipped partner (gray arrows)]. This degeneracy is lifted by interactions that mix the states, as shown in panel (i). The perturbation $\delta H_m$ conserves total spin $S$, which allows for a unique path to the LLL state at \mbox{$M=6$}, which is the Halperin~$(1,1,0)$ state shown schematically in (j), after another application of $\delta H_3$. 
    The LLL states at \mbox{$M=8$} are superpositions of the states (k)--(m). The excitation frequency should be fine-tuned to a particular final $S=0$ state.
    }  
    \label{fig:2}
\end{figure*}

Our starting point is the Hamiltonian of a two-component trapped Fermi gas with dimensionless contact-interaction strength $\tilde{g}$ in occupation number representation,
\begin{equation} \label{eq:hamiltonian}
H = \sum_{j,\sigma}\epsilon_{j}c^{\dagger}_{j\sigma}c_{j\sigma}^{} + \frac{\hbar^2}{m^*} \tilde{g} \sum_{ijkl}w_{ijkl}c^{\dagger}_{i\uparrow}c^{\dagger}_{j\downarrow}c_{k\downarrow}^{}c_{l\uparrow}^{} ,
\end{equation}
where we define the creation operator $c_{j\sigma}^{\dagger}$ of a fermion with mass $m^*$ and spin projection \mbox{$\sigma=\uparrow,\downarrow$} in a harmonic oscillator state \mbox{$j=\{n_j\geq0,k_j\geq0 \}$}. 
The single-particle energy in a rotating frame is~\cite{bloch08}
\begin{equation}\label{eq:spenergies}
\epsilon_j=\hbar \omega + \hbar(\omega+\Omega) n_j + \hbar(\omega-\Omega) k_j ,
\end{equation}
where $n_j$ is a Landau level index that denotes the number of cyclotron excitations and $k_j$ denotes the number of guiding-center excitations. The corresponding single-particle wave functions $\phi_j$ are independent of $\Omega$~\footnote{Explicitly, the single-particle wave function of a state \mbox{$j=\{n_j\geq 0,k_j \geq 0\}$} is $ \ell_{ho} \phi_j(z,\bar{z}) = \sqrt{\frac{\min(n_j,k_j)!}{\pi \max(n_j,k_j)!}} \ (z/\ell_{ho})^{|k_j-n_j|} e^{-\bar{z}z/2\ell_{ho}^2} \, L_{\min(n_j,k_j)}^{|k_j-n_j|}(\bar{z}z/\ell_{ho}^2)$, for $n_j\leq k_j $ and else $z$ exchanged with $\bar{z}$, where $\ell_{ho}$ is the harmonic oscillator length.}. We show this single-particle spectrum in Fig.~\ref{fig:1}, where states with equal Landau level index $n_j$ share a color and are labeled by angular momentum \mbox{$m_j = k_j - n_j \geq - n_j$} (the difference between the number of guiding-center and cyclotron excitations).
With increasing rotation frequency, the single-particle spectrum tilts, with energies shifted by an amount proportional to~$m_j$ [indicated by gray arrows in Fig.~\ref{fig:1}(a)]. In the frame of rapid rotation~\mbox{$\Omega \to \omega^{-}$} [Fig.~\ref{fig:1}(b)], Landau levels form in which states with different guiding-center excitations are degenerate~\cite{bloch08}.

Noninteracting ground and excited states of multiple particles are obtained by populating single-particle levels under a Pauli constraint. We illustrate this in Fig.~\ref{fig:1} for the same spin-balanced state of \mbox{$N=6$} fermions both in the stationary frame (where it forms the ground state) and in the frame of rapid rotations. 
Importantly, the Slater-determinant wave function of this state is independent of the rotation frequency, and excitations to any state with equal total angular momentum are gapped by multiples of \mbox{$\hbar\omega_c = 2 \hbar\omega$} [since by Eq.~\eqref{eq:spenergies} the energy shift between two states due to rotations is $\Delta E = - \Delta M \hbar \Omega$, where $\Delta M$ is the difference in total angular momentum].
The noninteracting degeneracy is lifted by the interaction matrix element in Eq.~\eqref{eq:hamiltonian}, which is set by the overlap integral for the contact interaction,
\begin{align} \label{eq:overlap}
w_{ijkl} = \int d^2r \, \phi_i^{*}\phi_j^{*}\phi_k\phi_l .
\end{align}
The dimensionless coupling constant \mbox{$\tilde{g}$} in Eq.~\eqref{eq:hamiltonian} is linked to the three-dimensional $s$-wave scattering length by \mbox{$\tilde{g} = \sqrt{8\pi}a_{3D}/\ell_z$}, where $\ell_z$ is the oscillator length of a transverse confining potential~\cite{bloch08,zwerger16,hofmann21}.
For single-particle wave functions in the LLL, this contact interaction is then described by a Haldane pseudopotential \mbox{$V_0/\hbar\omega = \tilde{g}/2\pi$}~\cite{Haldane83}, and $w_{ijkl}$ involve combinatorial factors~\cite{craps24}.
First-order degenerate perturbation theory predicts a shift, linear in the coupling constant, from the noninteracting level energy. This shift is obtained by exact diagonalization of the Hamiltonian~\eqref{eq:hamiltonian} restricted to the subspace of states that are degenerate without interactions and have equal total angular momentum~\cite{sakurai94}. Since, by definition, the noninteracting energy is diagonal and constant within the subspace, the interaction matrix is proportional to the interaction strength. Diagonalization then gives an $\it{O}(\tilde{g})$ correction to the noninteracting energy, with eigenstates that are themselves independent of the coupling. Perturbation theory is applicable provided that the splitting is small compared to the noninteracting gap $2\hbar\omega$. In particular, degenerate perturbation theory in a nonrotating (or rotating) trap is then equivalent to exact diagonalization in the LLL, where the requirement to stay within the LLL is that the energy splitting due to the interaction be small compared to $\hbar\omega_c=2\hbar\omega$~\cite{footnote}. This shows that FQH states are realized as excited states even in nonrotating traps.
\begin{figure}[t!]
    { 
{\includegraphics{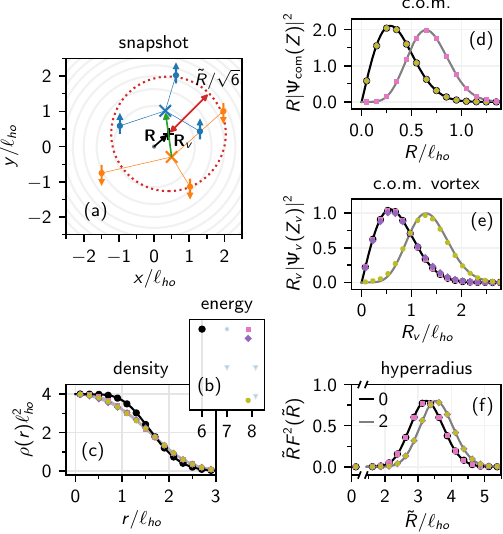}}}
    \caption{(a) Snapshot of spin-resolved $\sigma=\uparrow,\downarrow$ particle positions (blue and orange points), where \mbox{$\ell_{ho}=\sqrt{\hbar/(m^{*}\omega)}$} is the oscillator length, obtained by Monte Carlo sampling of the many-body wave function of the Halperin~$(1,1,0)$ state. The harmonic trap is indicated by gray contour lines. Blue (orange) crosses indicate the c.o.m.~coordinate for spin projection \mbox{$\sigma=\uparrow(\downarrow)$}. Black, green, and red arrows show the c.o.m.~vector~${\bf R}$, the c.o.m.~vortex vector~${\bf R}_v$, and the rescaled hyperradius \mbox{$\tilde{R}/\sqrt{N}$}, respectively. (b) LLL states from Fig.~\ref{fig:2}(n), but with an individual visual coding for the set of states at \mbox{$M=6,8$} that are excitable by $\delta H_m$ [Eq.~\eqref{eq:excitation}]. 
    With the extended histograms in (d)--(f), each state can be uniquely distinguished. }    
    \label{fig:3}
\end{figure}

In Fig.~\ref{fig:2}, we illustrate the key point that quantum Hall states are realized as excited states in a harmonic trap at weak interactions for~\mbox{$N=6$} spin-balanced fermions. The figure shows the evolution of the excitation spectrum with rotation frequencies \mbox{$\Omega=0,\omega/3,$} and~$\omega$~[(a)--(c)]. We choose a perturbative interaction strength~\mbox{$\tilde{g}=1$}, which is large enough to split the spectrum while states remain visibly clustered around their noninteracting energy. 
We color states according to their highest occupied single-particle Landau level [cf.~Fig.~\ref{fig:1}], where states within the LLL are blue. 
The noninteracting level splitting of \mbox{$\hbar\omega_c=2\hbar\omega$} between clusters of states within a given total angular momentum sector $M$ is clearly visible. Moreover, the relative splitting of states within each cluster is unchanged across \mbox{(a)--(c)}, as are their wave functions, and the only effect of rotations is to shift energies by \mbox{$\Delta E = - M \hbar \Omega$}. Orange- and blue-shaded regions highlight this rearrangement for a set of states, the latter being LLL states with minimal $M$.
Panel~(c) shows the conventional representation of the LLL, within which, as discussed, the contact interaction is described by a single pseudopotential $V_0$. Such a setup realizes exact Halperin states~\cite{trugman85}, and we mark the $(1,1,0)$ Halperin state at \mbox{$M=6$} (\mbox{$\nu = 2$}) and the $(1,1,1)$ state at \mbox{$M=15$} (\mbox{$\nu = 1$}), the latter being the minimal-$M$ LLL state with vanishing interaction energy. At higher angular momenta, many additional FQH states are realized, including skyrmion states~\cite{palm20}. All of these states are already present in the stationary trap as excited states [Fig.~\ref{fig:2}(a)].

For further illustration, the bottom row of Fig.~\ref{fig:2} shows the occupation of the ground state [circle in (a)] as well as the orange- and blue-shaded states. We show these states in the rapidly rotating frame [cf.~Fig.~\ref{fig:1}], which, as discussed, has no bearing on the many-body wave function itself. Panels (i) and (n) show a more detailed view of the spectrum that also highlights the total spin of states, where the spin~\mbox{$S=0$} (\mbox{$S=1$}) states form a singlet (triplet) of the spins in unpaired levels. 
We also indicate a possible excitation pathway from the ground state to the first states in the LLL (with \mbox{$M=6,8$}) using the perturbation~\eqref{eq:excitation} by black and gray arrows, respectively. Such an excitation can proceed by a successive transfer of states with $\delta H_{m=3(4)}$ via the highlighted states at \mbox{$M=3(4)$}. Since the nonrotating ground state in (d) is a total spin singlet and $\delta H_m$ conserves spin, the marked \mbox{$M=3$} state can be uniquely realized in this scheme, while reaching a particular state at \mbox{$M=4$} requires fine-tuning the excitation frequency. A second application of \mbox{$\delta H_{3(4)}$} then transfers the remaining particle in the cyclotron excitation to the LLL state shown in (j) [(k)], with (j)~corresponding to the \mbox{$(1,1,0)$} Halperin state, which is realized uniquely in this scheme. To realize a particular \mbox{$M=8$} LLL state, which is a superposition of~(k) and the two singlet configurations in (l) and (m), the excitation frequency should be fine-tuned with a precision smaller than the energy splitting in~(n), which is well within the reach of current experiments as demonstrated in Refs.~\cite{bayha20,lunt24}.

Given the ability to experimentally sample the many-body wave function using spin-resolved single-particle imaging, the present setup should not only act as a quantum simulator of LLL states but also provide novel means to probe their many-body structure. We conclude by illustrating how such imaging can be used to distinguish the four \mbox{$S=0$} LLL states at \mbox{$M=6,8$} discussed above. Such experiments proceed by taking a snapshot of particle positions using fluorescence imaging~\cite{bergschneider18}, or by first pinning atoms onto an optical lattice~\cite{Xiang25,Yao25,deJongh25}, for every realization of a final state.
Figure~\ref{fig:3}(a) shows a single snapshot obtained by Monte Carlo sampling of an eigenstate, where blue (orange) points indicate spin-$\uparrow$ (spin-$\downarrow$) positions and the harmonic trap is indicated by gray contour lines. Repeating this procedure many times gives statistical information on the many-body wave function. A natural observable for the four LLL states [highlighted in Fig.~\ref{fig:3}(b)] is the particle density. We show corresponding results in Fig.~\ref{fig:3}(c), where solid points are results from Metropolis Monte Carlo sampling of the many-body wave functions obtained from diagonalization as described above, and the curves are analytical predictions. As is apparent from the figure, however, different states are hardly distinguishable, calling for different observables. A possible observable is the spin-resolved pair correlation function, which contains information about relative angular momentum pairing, and which can be extracted by sampling interparticle distances. These pair correlation functions are sensitive to interaction effects at smaller filling factors (larger $M$), but for the states considered here ($\nu \lesssim  2$) take a similar shape~\cite{footnote}.
Microscopically, we find that the highest-energy LLL state at~\mbox{$M=8$} [pink square] is a c.o.m. excitation of the $(1,1,0)$ Halperin state [black circle], i.e., \mbox{$\sim\Psi_{(1,1,0)}Z^2$}, where \mbox{$Z=\frac{1}{N}\sum_{i,\sigma} z_{i\sigma}$} is the c.o.m.~coordinate in the complex plane. Likewise, the lowest-energy state at~\mbox{$M=8$} [yellow octagon] is almost exactly (with $96\%$ weight, becoming exact in the thermodynamic limit)
a c.o.m.~vortex excitation \mbox{$\sim\Psi_{(1,1,0)}Z_v^2$} with c.o.m.~vortex coordinate \mbox{$Z_v = Z_\uparrow - Z_\downarrow$}, where $Z_\sigma$ is the c.o.m.~coordinate of the spin components. Finally, all states have a hyperradial degree of freedom, parametrized by the coordinate~\mbox{$\tilde{R} = \sqrt{\sum_{i,\sigma} \left| \mathbf{r}_{i\sigma}- \mathbf{R} \right|^2}$}, where both the c.o.m.~vortex excitation and the intermediate-energy \mbox{$M=8$} state [purple diamond] differ from the other two (formally, they are derived from primary states at distinct noninteracting energy levels~\cite{bekassy22,bekassy24}). These three degrees of freedom are easily accessible from single-shot images, and we indicate the c.o.m.~position, the c.o.m.~vortex, and the hyperradius by black, green, and red arrows in Fig.~\ref{fig:3}(a), respectively. Figures~\ref{fig:3}(d)--\ref{fig:3}(f) show the resulting distributions of (d)~the c.o.m.~coordinate \mbox{$R=|Z|$}, (e)~the c.o.m.~vortex coordinate \mbox{$R_v=|Z_v|$}, and (f)~the hyperradial coordinate~$\tilde{R}$. As before, markers are the results of Monte Carlo sampling, and black and gray continuous lines indicate analytical distributions. Taken together, all four states are easily distinguishable by inspection of (d)--(f), where states with c.o.m., c.o.m. vortex, or hyperradial excitations have distributions that are pushed to larger parameter values. 

In summary, building on the established correspondence between LLL states at rapid rotation and excited states of a weakly interacting Fermi gas at fixed total angular momentum~\cite{wilkin98,cooper08}, we have shown how the excitation protocol of \textcite{lunt24} can be used to reach targeted many-body LLL states beyond two particles, and we have proposed how single-particle imaging can distinguish the resulting states.
The identification holds under the condition that the interaction strength is in the perturbative regime with respect to the harmonic oscillator spacing. 
While trap anharmonicity and, in particular, trap anisotropy (which will mix different Landau levels) might limit the number of particles in experiments~\cite{footnote}, a distinct advantage of such setups is that general excited states within the LLL are accessible. An interesting avenue for further study would be to devise more complex excitation protocols that prepare excited states in the LLL, such as the magnetoroton~\cite{girvin86,kukushkin09}, graviton excitations~\cite{haldane11,liou19,pinczuk93,liang24}, Wigner droplets~\cite{wigner34,Saarikoski10,reimann02}, or anyon excitations~\cite{Leinaas77,goldin80,wilczek82,arovas84}. Finally, we note that the discussion above is not restricted to two-component systems, but applies equally well to spin-polarized fermions (which are described by a single pseudopotential parameter \mbox{$V_1$} in the LLL~\cite{Regnault04,bloch08} with exact Laughlin wave functions as eigenstates~\cite{trugman85}) or dipolar fermions (with two dominant pseudopotentials~$V_1$ and~$V_3$, or $V_0$ and $V_2$ for two-component fermions~\cite{cooper05}). 

\begin{acknowledgments}
\textit{Acknowledgments}---We thank N. Cooper, P. Lunt, and W. Zwerger for discussions. 
This work is supported by Chalmers' Nano Area of Advance (Excellence grant), Vetenskapsr\aa det (Grants No.~2020-04239 and No. 2024-04485), the Olle Engkvist Foundation (Grant No.~233-0339), the Knut and Alice Wallenberg Foundation (Grant No.~KAW 2024.0129), and Nordita.
\end{acknowledgments}

\textit{Data availability}---There are no publicly available research data or software supporting this manuscript. Requests for further information or data should be sent to the authors.
\bibliography{bib_fqhe}

\begin{thebibliography}{66}%
\makeatletter
\providecommand \@ifxundefined [1]{%
 \@ifx{#1\undefined}
}%
\providecommand \@ifnum [1]{%
 \ifnum #1\expandafter \@firstoftwo
 \else \expandafter \@secondoftwo
 \fi
}%
\providecommand \@ifx [1]{%
 \ifx #1\expandafter \@firstoftwo
 \else \expandafter \@secondoftwo
 \fi
}%
\providecommand \natexlab [1]{#1}%
\providecommand \enquote  [1]{``#1''}%
\providecommand \bibnamefont  [1]{#1}%
\providecommand \bibfnamefont [1]{#1}%
\providecommand \citenamefont [1]{#1}%
\providecommand \href@noop [0]{\@secondoftwo}%
\providecommand \href [0]{\begingroup \@sanitize@url \@href}%
\providecommand \@href[1]{\@@startlink{#1}\@@href}%
\providecommand \@@href[1]{\endgroup#1\@@endlink}%
\providecommand \@sanitize@url [0]{\catcode `\\12\catcode `\$12\catcode `\&12\catcode `\#12\catcode `\^12\catcode `\_12\catcode `\%12\relax}%
\providecommand \@@startlink[1]{}%
\providecommand \@@endlink[0]{}%
\providecommand \url  [0]{\begingroup\@sanitize@url \@url }%
\providecommand \@url [1]{\endgroup\@href {#1}{\urlprefix }}%
\providecommand \urlprefix  [0]{URL }%
\providecommand \Eprint [0]{\href }%
\providecommand \doibase [0]{https://doi.org/}%
\providecommand \selectlanguage [0]{\@gobble}%
\providecommand \bibinfo  [0]{\@secondoftwo}%
\providecommand \bibfield  [0]{\@secondoftwo}%
\providecommand \translation [1]{[#1]}%
\providecommand \BibitemOpen [0]{}%
\providecommand \bibitemStop [0]{}%
\providecommand \bibitemNoStop [0]{.\EOS\space}%
\providecommand \EOS [0]{\spacefactor3000\relax}%
\providecommand \BibitemShut  [1]{\csname bibitem#1\endcsname}%
\let\auto@bib@innerbib\@empty
\bibitem [{\citenamefont {Anderson}\ \emph {et~al.}(1995)\citenamefont {Anderson}, \citenamefont {Ensher}, \citenamefont {Matthews}, \citenamefont {Wieman},\ and\ \citenamefont {Cornell}}]{Anderson95}%
  \BibitemOpen
  \bibfield  {author} {\bibinfo {author} {\bibfnamefont {M.~H.}\ \bibnamefont {Anderson}}, \bibinfo {author} {\bibfnamefont {J.~R.}\ \bibnamefont {Ensher}}, \bibinfo {author} {\bibfnamefont {M.~R.}\ \bibnamefont {Matthews}}, \bibinfo {author} {\bibfnamefont {C.~E.}\ \bibnamefont {Wieman}},\ and\ \bibinfo {author} {\bibfnamefont {E.~A.}\ \bibnamefont {Cornell}},\ }\bibfield  {title} {\bibinfo {title} {{Observation of Bose-Einstein condensation in a dilute atomic vapor}},\ }\href {https://doi.org/10.1126/science.269.5221.198} {\bibfield  {journal} {\bibinfo  {journal} {Science}\ }\textbf {\bibinfo {volume} {269}},\ \bibinfo {pages} {198} (\bibinfo {year} {1995})}\BibitemShut {NoStop}%
\bibitem [{\citenamefont {Davis}\ \emph {et~al.}(1995)\citenamefont {Davis}, \citenamefont {Mewes}, \citenamefont {Andrews}, \citenamefont {van Druten}, \citenamefont {Durfee}, \citenamefont {Kurn},\ and\ \citenamefont {Ketterle}}]{davis95}%
  \BibitemOpen
  \bibfield  {author} {\bibinfo {author} {\bibfnamefont {K.~B.}\ \bibnamefont {Davis}}, \bibinfo {author} {\bibfnamefont {M.~O.}\ \bibnamefont {Mewes}}, \bibinfo {author} {\bibfnamefont {M.~R.}\ \bibnamefont {Andrews}}, \bibinfo {author} {\bibfnamefont {N.~J.}\ \bibnamefont {van Druten}}, \bibinfo {author} {\bibfnamefont {D.~S.}\ \bibnamefont {Durfee}}, \bibinfo {author} {\bibfnamefont {D.~M.}\ \bibnamefont {Kurn}},\ and\ \bibinfo {author} {\bibfnamefont {W.}~\bibnamefont {Ketterle}},\ }\bibfield  {title} {\bibinfo {title} {{Bose-Einstein Condensation in a Gas of Sodium Atoms}},\ }\href {https://doi.org/10.1103/PhysRevLett.75.3969} {\bibfield  {journal} {\bibinfo  {journal} {Phys. Rev. Lett.}\ }\textbf {\bibinfo {volume} {75}},\ \bibinfo {pages} {3969} (\bibinfo {year} {1995})}\BibitemShut {NoStop}%
\bibitem [{\citenamefont {Cooper}(2008)}]{cooper08}%
  \BibitemOpen
  \bibfield  {author} {\bibinfo {author} {\bibfnamefont {N.~R.}\ \bibnamefont {Cooper}},\ }\bibfield  {title} {\bibinfo {title} {{Rapidly rotating atomic gases}},\ }\href {https://doi.org/10.1080/00018730802564122} {\bibfield  {journal} {\bibinfo  {journal} {Adv. Phys.}\ }\textbf {\bibinfo {volume} {57}},\ \bibinfo {pages} {539} (\bibinfo {year} {2008})}\BibitemShut {NoStop}%
\bibitem [{\citenamefont {Bloch}\ \emph {et~al.}(2008)\citenamefont {Bloch}, \citenamefont {Dalibard},\ and\ \citenamefont {Zwerger}}]{bloch08}%
  \BibitemOpen
  \bibfield  {author} {\bibinfo {author} {\bibfnamefont {I.}~\bibnamefont {Bloch}}, \bibinfo {author} {\bibfnamefont {J.}~\bibnamefont {Dalibard}},\ and\ \bibinfo {author} {\bibfnamefont {W.}~\bibnamefont {Zwerger}},\ }\bibfield  {title} {\bibinfo {title} {{Many-body physics with ultracold gases}},\ }\href {https://doi.org/10.1103/RevModPhys.80.885} {\bibfield  {journal} {\bibinfo  {journal} {Rev. Mod. Phys.}\ }\textbf {\bibinfo {volume} {80}},\ \bibinfo {pages} {885} (\bibinfo {year} {2008})}\BibitemShut {NoStop}%
\bibitem [{\citenamefont {Madison}\ \emph {et~al.}(2000)\citenamefont {Madison}, \citenamefont {Chevy}, \citenamefont {Wohlleben},\ and\ \citenamefont {Dalibard}}]{madison00}%
  \BibitemOpen
  \bibfield  {author} {\bibinfo {author} {\bibfnamefont {K.~W.}\ \bibnamefont {Madison}}, \bibinfo {author} {\bibfnamefont {F.}~\bibnamefont {Chevy}}, \bibinfo {author} {\bibfnamefont {W.}~\bibnamefont {Wohlleben}},\ and\ \bibinfo {author} {\bibfnamefont {J.}~\bibnamefont {Dalibard}},\ }\bibfield  {title} {\bibinfo {title} {{Vortex Formation in a Stirred Bose-Einstein Condensate}},\ }\href {https://doi.org/10.1103/PhysRevLett.84.806} {\bibfield  {journal} {\bibinfo  {journal} {Phys. Rev. Lett.}\ }\textbf {\bibinfo {volume} {84}},\ \bibinfo {pages} {806} (\bibinfo {year} {2000})}\BibitemShut {NoStop}%
\bibitem [{\citenamefont {Abo-Shaeer}\ \emph {et~al.}(2001)\citenamefont {Abo-Shaeer}, \citenamefont {Raman}, \citenamefont {Vogels},\ and\ \citenamefont {Ketterle}}]{Shaeer01}%
  \BibitemOpen
  \bibfield  {author} {\bibinfo {author} {\bibfnamefont {J.~R.}\ \bibnamefont {Abo-Shaeer}}, \bibinfo {author} {\bibfnamefont {J.}~\bibnamefont {Raman}}, \bibinfo {author} {\bibfnamefont {J.~M.}\ \bibnamefont {Vogels}},\ and\ \bibinfo {author} {\bibfnamefont {W.}~\bibnamefont {Ketterle}},\ }\bibfield  {title} {\bibinfo {title} {{Observation of Vortex Lattices in Bose-Einstein Condensates}},\ }\href {https://doi.org/10.1126/science.1060182} {\bibfield  {journal} {\bibinfo  {journal} {Science}\ }\textbf {\bibinfo {volume} {292}},\ \bibinfo {pages} {476} (\bibinfo {year} {2001})}\BibitemShut {NoStop}%
\bibitem [{\citenamefont {Cr\'epel}\ \emph {et~al.}(2023)\citenamefont {Cr\'epel}, \citenamefont {Yao}, \citenamefont {Mukherjee}, \citenamefont {Fletcher},\ and\ \citenamefont {Zwierlein}}]{crepel23}%
  \BibitemOpen
  \bibfield  {author} {\bibinfo {author} {\bibfnamefont {V.}~\bibnamefont {Cr\'epel}}, \bibinfo {author} {\bibfnamefont {R.}~\bibnamefont {Yao}}, \bibinfo {author} {\bibfnamefont {B.}~\bibnamefont {Mukherjee}}, \bibinfo {author} {\bibfnamefont {R.}~\bibnamefont {Fletcher}},\ and\ \bibinfo {author} {\bibfnamefont {M.}~\bibnamefont {Zwierlein}},\ }\bibfield  {title} {\bibinfo {title} {{Geometric squeezing of rotating quantum gases into the lowest {Landau} level}},\ }\href {https://doi.org/10.5802/crphys.173} {\bibfield  {journal} {\bibinfo  {journal} {C. R. Phys.}\ }\textbf {\bibinfo {volume} {24}},\ \bibinfo {pages} {241} (\bibinfo {year} {2023})}\BibitemShut {NoStop}%
\bibitem [{\citenamefont {Fletcher}\ \emph {et~al.}(2021)\citenamefont {Fletcher}, \citenamefont {Shaffer}, \citenamefont {Wilson}, \citenamefont {Patel}, \citenamefont {Yan}, \citenamefont {Cr{\'e}pel}, \citenamefont {Mukherjee},\ and\ \citenamefont {Zwierlein}}]{fletcher21}%
  \BibitemOpen
  \bibfield  {author} {\bibinfo {author} {\bibfnamefont {R.~J.}\ \bibnamefont {Fletcher}}, \bibinfo {author} {\bibfnamefont {A.}~\bibnamefont {Shaffer}}, \bibinfo {author} {\bibfnamefont {C.~C.}\ \bibnamefont {Wilson}}, \bibinfo {author} {\bibfnamefont {P.~B.}\ \bibnamefont {Patel}}, \bibinfo {author} {\bibfnamefont {Z.}~\bibnamefont {Yan}}, \bibinfo {author} {\bibfnamefont {V.}~\bibnamefont {Cr{\'e}pel}}, \bibinfo {author} {\bibfnamefont {B.}~\bibnamefont {Mukherjee}},\ and\ \bibinfo {author} {\bibfnamefont {M.~W.}\ \bibnamefont {Zwierlein}},\ }\bibfield  {title} {\bibinfo {title} {Geometric squeezing into the lowest {Landau} level},\ }\href {https://doi.org/10.1126/science.aba7202} {\bibfield  {journal} {\bibinfo  {journal} {Science}\ }\textbf {\bibinfo {volume} {372}},\ \bibinfo {pages} {1318} (\bibinfo {year} {2021})}\BibitemShut {NoStop}%
\bibitem [{\citenamefont {Mukherjee}\ \emph {et~al.}(2022)\citenamefont {Mukherjee}, \citenamefont {Shaffer}, \citenamefont {Patel}, \citenamefont {Yan}, \citenamefont {Wilson}, \citenamefont {Cr{\'e}pel}, \citenamefont {Fletcher},\ and\ \citenamefont {Zwierlein}}]{mukherjee22}%
  \BibitemOpen
  \bibfield  {author} {\bibinfo {author} {\bibfnamefont {B.}~\bibnamefont {Mukherjee}}, \bibinfo {author} {\bibfnamefont {A.}~\bibnamefont {Shaffer}}, \bibinfo {author} {\bibfnamefont {P.~B.}\ \bibnamefont {Patel}}, \bibinfo {author} {\bibfnamefont {Z.}~\bibnamefont {Yan}}, \bibinfo {author} {\bibfnamefont {C.~C.}\ \bibnamefont {Wilson}}, \bibinfo {author} {\bibfnamefont {V.}~\bibnamefont {Cr{\'e}pel}}, \bibinfo {author} {\bibfnamefont {R.~J.}\ \bibnamefont {Fletcher}},\ and\ \bibinfo {author} {\bibfnamefont {M.}~\bibnamefont {Zwierlein}},\ }\bibfield  {title} {\bibinfo {title} {{Crystallization of bosonic quantum Hall states in a rotating quantum gas}},\ }\href {https://doi.org/10.1038/s41586-021-04170-2} {\bibfield  {journal} {\bibinfo  {journal} {Nature (London)}\ }\textbf {\bibinfo {volume} {601}},\ \bibinfo {pages} {58} (\bibinfo {year} {2022})}\BibitemShut {NoStop}%
\bibitem [{\citenamefont {Viefers}(2008)}]{viefers08}%
  \BibitemOpen
  \bibfield  {author} {\bibinfo {author} {\bibfnamefont {S.}~\bibnamefont {Viefers}},\ }\bibfield  {title} {\bibinfo {title} {{Quantum Hall physics in rotating Bose–Einstein condensates}},\ }\href {https://doi.org/10.1088/0953-8984/20/12/123202} {\bibfield  {journal} {\bibinfo  {journal} {J. Phys. Condens. Matter}\ }\textbf {\bibinfo {volume} {20}},\ \bibinfo {pages} {123202} (\bibinfo {year} {2008})}\BibitemShut {NoStop}%
\bibitem [{\citenamefont {Dalibard}\ \emph {et~al.}(2011)\citenamefont {Dalibard}, \citenamefont {Gerbier}, \citenamefont {Juzeli\ifmmode~\bar{u}\else \={u}\fi{}nas},\ and\ \citenamefont {\"Ohberg}}]{dalibard11}%
  \BibitemOpen
  \bibfield  {author} {\bibinfo {author} {\bibfnamefont {J.}~\bibnamefont {Dalibard}}, \bibinfo {author} {\bibfnamefont {F.}~\bibnamefont {Gerbier}}, \bibinfo {author} {\bibfnamefont {G.}~\bibnamefont {Juzeli\ifmmode~\bar{u}\else \={u}\fi{}nas}},\ and\ \bibinfo {author} {\bibfnamefont {P.}~\bibnamefont {\"Ohberg}},\ }\bibfield  {title} {\bibinfo {title} {{Colloquium: Artificial gauge potentials for neutral atoms}},\ }\href {https://doi.org/10.1103/RevModPhys.83.1523} {\bibfield  {journal} {\bibinfo  {journal} {Rev. Mod. Phys.}\ }\textbf {\bibinfo {volume} {83}},\ \bibinfo {pages} {1523} (\bibinfo {year} {2011})}\BibitemShut {NoStop}%
\bibitem [{\citenamefont {Dalibard}()}]{dalibard15}%
  \BibitemOpen
  \bibfield  {author} {\bibinfo {author} {\bibfnamefont {J.}~\bibnamefont {Dalibard}},\ }\href {https://arxiv.org/abs/1504.05520} {\bibinfo {title} {{Introduction to the physics of artificial gauge fields}}},\ \Eprint {https://arxiv.org/abs/1504.05520} {arXiv:1504.05520} \BibitemShut {NoStop}%
\bibitem [{\citenamefont {Aidelsburger}\ \emph {et~al.}(2011)\citenamefont {Aidelsburger}, \citenamefont {Atala}, \citenamefont {Nascimb\`ene}, \citenamefont {Trotzky}, \citenamefont {Chen},\ and\ \citenamefont {Bloch}}]{aidelsburger11}%
  \BibitemOpen
  \bibfield  {author} {\bibinfo {author} {\bibfnamefont {M.}~\bibnamefont {Aidelsburger}}, \bibinfo {author} {\bibfnamefont {M.}~\bibnamefont {Atala}}, \bibinfo {author} {\bibfnamefont {S.}~\bibnamefont {Nascimb\`ene}}, \bibinfo {author} {\bibfnamefont {S.}~\bibnamefont {Trotzky}}, \bibinfo {author} {\bibfnamefont {Y.-A.}\ \bibnamefont {Chen}},\ and\ \bibinfo {author} {\bibfnamefont {I.}~\bibnamefont {Bloch}},\ }\bibfield  {title} {\bibinfo {title} {{Experimental Realization of Strong Effective Magnetic Fields in an Optical Lattice}},\ }\href {https://doi.org/10.1103/PhysRevLett.107.255301} {\bibfield  {journal} {\bibinfo  {journal} {Phys. Rev. Lett.}\ }\textbf {\bibinfo {volume} {107}},\ \bibinfo {pages} {255301} (\bibinfo {year} {2011})}\BibitemShut {NoStop}%
\bibitem [{\citenamefont {Struck}\ \emph {et~al.}(2012)\citenamefont {Struck}, \citenamefont {\"Olschl\"ager}, \citenamefont {Weinberg}, \citenamefont {Hauke}, \citenamefont {Simonet}, \citenamefont {Eckardt}, \citenamefont {Lewenstein}, \citenamefont {Sengstock},\ and\ \citenamefont {Windpassinger}}]{struck12}%
  \BibitemOpen
  \bibfield  {author} {\bibinfo {author} {\bibfnamefont {J.}~\bibnamefont {Struck}}, \bibinfo {author} {\bibfnamefont {C.}~\bibnamefont {\"Olschl\"ager}}, \bibinfo {author} {\bibfnamefont {M.}~\bibnamefont {Weinberg}}, \bibinfo {author} {\bibfnamefont {P.}~\bibnamefont {Hauke}}, \bibinfo {author} {\bibfnamefont {J.}~\bibnamefont {Simonet}}, \bibinfo {author} {\bibfnamefont {A.}~\bibnamefont {Eckardt}}, \bibinfo {author} {\bibfnamefont {M.}~\bibnamefont {Lewenstein}}, \bibinfo {author} {\bibfnamefont {K.}~\bibnamefont {Sengstock}},\ and\ \bibinfo {author} {\bibfnamefont {P.}~\bibnamefont {Windpassinger}},\ }\bibfield  {title} {\bibinfo {title} {{Tunable Gauge Potential for Neutral and Spinless Particles in Driven Optical Lattices}},\ }\href {https://doi.org/10.1103/PhysRevLett.108.225304} {\bibfield  {journal} {\bibinfo  {journal} {Phys. Rev. Lett.}\ }\textbf {\bibinfo {volume} {108}},\ \bibinfo {pages} {225304} (\bibinfo {year} {2012})}\BibitemShut {NoStop}%
\bibitem [{\citenamefont {Lin}\ \emph {et~al.}(2009)\citenamefont {Lin}, \citenamefont {Compton}, \citenamefont {Jiménez-García}, \citenamefont {Porto},\ and\ \citenamefont {Spielman}}]{Lin09}%
  \BibitemOpen
  \bibfield  {author} {\bibinfo {author} {\bibfnamefont {Y.}~\bibnamefont {Lin}}, \bibinfo {author} {\bibfnamefont {R.}~\bibnamefont {Compton}}, \bibinfo {author} {\bibfnamefont {K.}~\bibnamefont {Jiménez-García}}, \bibinfo {author} {\bibfnamefont {J.~V.}\ \bibnamefont {Porto}},\ and\ \bibinfo {author} {\bibfnamefont {I.~B.}\ \bibnamefont {Spielman}},\ }\bibfield  {title} {\bibinfo {title} {{Synthetic magnetic fields for ultracold neutral atoms}},\ }\href {https://doi.org/10.1038/nature08609} {\bibfield  {journal} {\bibinfo  {journal} {Nature (London)}\ }\textbf {\bibinfo {volume} {462}},\ \bibinfo {pages} {628} (\bibinfo {year} {2009})}\BibitemShut {NoStop}%
\bibitem [{\citenamefont {Nascimbene}(2025)}]{nascimbene25}%
  \BibitemOpen
  \bibfield  {author} {\bibinfo {author} {\bibfnamefont {S.}~\bibnamefont {Nascimbene}},\ }\bibfield  {title} {\bibinfo {title} {{Simulating quantum {Hall} physics in ultracold atomic gases: Prospects and challenges}},\ }\href {https://doi.org/10.5802/crphys.243} {\bibfield  {journal} {\bibinfo  {journal} {C. R. Phys.}\ }\textbf {\bibinfo {volume} {26}},\ \bibinfo {pages} {317} (\bibinfo {year} {2025})}\BibitemShut {NoStop}%
\bibitem [{\citenamefont {Gemelke}\ \emph {et~al.}()\citenamefont {Gemelke}, \citenamefont {Sarajlic},\ and\ \citenamefont {Chu}}]{gemelke10}%
  \BibitemOpen
  \bibfield  {author} {\bibinfo {author} {\bibfnamefont {N.}~\bibnamefont {Gemelke}}, \bibinfo {author} {\bibfnamefont {E.}~\bibnamefont {Sarajlic}},\ and\ \bibinfo {author} {\bibfnamefont {S.}~\bibnamefont {Chu}},\ }\href {https://arxiv.org/abs/1007.2677} {\bibinfo {title} {{Rotating few-body atomic systems in the fractional quantum Hall regime}}},\ \Eprint {https://arxiv.org/abs/1007.2677} {arXiv:1007.2677} \BibitemShut {NoStop}%
\bibitem [{\citenamefont {Clark}\ \emph {et~al.}(2020)\citenamefont {Clark}, \citenamefont {Schine}, \citenamefont {Baum}, \citenamefont {Jia},\ and\ \citenamefont {Simon}}]{clark20}%
  \BibitemOpen
  \bibfield  {author} {\bibinfo {author} {\bibfnamefont {L.~W.}\ \bibnamefont {Clark}}, \bibinfo {author} {\bibfnamefont {N.}~\bibnamefont {Schine}}, \bibinfo {author} {\bibfnamefont {C.}~\bibnamefont {Baum}}, \bibinfo {author} {\bibfnamefont {N.}~\bibnamefont {Jia}},\ and\ \bibinfo {author} {\bibfnamefont {J.}~\bibnamefont {Simon}},\ }\bibfield  {title} {\bibinfo {title} {{Observation of Laughlin states made of light}},\ }\href {https://doi.org/10.1038/s41586-020-2318-5} {\bibfield  {journal} {\bibinfo  {journal} {Nature (London)}\ }\textbf {\bibinfo {volume} {582}},\ \bibinfo {pages} {41} (\bibinfo {year} {2020})}\BibitemShut {NoStop}%
\bibitem [{\citenamefont {L{\'e}onard}\ \emph {et~al.}(2023)\citenamefont {L{\'e}onard}, \citenamefont {Kim}, \citenamefont {Kwan}, \citenamefont {Segura}, \citenamefont {Grusdt}, \citenamefont {Repellin}, \citenamefont {Goldman},\ and\ \citenamefont {Greiner}}]{leonard23}%
  \BibitemOpen
  \bibfield  {author} {\bibinfo {author} {\bibfnamefont {J.}~\bibnamefont {L{\'e}onard}}, \bibinfo {author} {\bibfnamefont {S.}~\bibnamefont {Kim}}, \bibinfo {author} {\bibfnamefont {J.}~\bibnamefont {Kwan}}, \bibinfo {author} {\bibfnamefont {P.}~\bibnamefont {Segura}}, \bibinfo {author} {\bibfnamefont {F.}~\bibnamefont {Grusdt}}, \bibinfo {author} {\bibfnamefont {C.}~\bibnamefont {Repellin}}, \bibinfo {author} {\bibfnamefont {N.}~\bibnamefont {Goldman}},\ and\ \bibinfo {author} {\bibfnamefont {M.}~\bibnamefont {Greiner}},\ }\bibfield  {title} {\bibinfo {title} {{Realization of a fractional quantum Hall state with ultracold atoms}},\ }\href {https://doi.org/10.1038/s41586-023-06122-4} {\bibfield  {journal} {\bibinfo  {journal} {Nature (London)}\ }\textbf {\bibinfo {volume} {619}},\ \bibinfo {pages} {495} (\bibinfo {year} {2023})}\BibitemShut {NoStop}%
\bibitem [{\citenamefont {Lunt}\ \emph {et~al.}(2024{\natexlab{a}})\citenamefont {Lunt}, \citenamefont {Hill}, \citenamefont {Reiter}, \citenamefont {Preiss}, \citenamefont {Ga\l{}ka},\ and\ \citenamefont {Jochim}}]{lunt24}%
  \BibitemOpen
  \bibfield  {author} {\bibinfo {author} {\bibfnamefont {P.}~\bibnamefont {Lunt}}, \bibinfo {author} {\bibfnamefont {P.}~\bibnamefont {Hill}}, \bibinfo {author} {\bibfnamefont {J.}~\bibnamefont {Reiter}}, \bibinfo {author} {\bibfnamefont {P.~M.}\ \bibnamefont {Preiss}}, \bibinfo {author} {\bibfnamefont {M.}~\bibnamefont {Ga\l{}ka}},\ and\ \bibinfo {author} {\bibfnamefont {S.}~\bibnamefont {Jochim}},\ }\bibfield  {title} {\bibinfo {title} {{Realization of a Laughlin State of Two Rapidly Rotating Fermions}},\ }\href {https://doi.org/10.1103/PhysRevLett.133.253401} {\bibfield  {journal} {\bibinfo  {journal} {Phys. Rev. Lett.}\ }\textbf {\bibinfo {volume} {133}},\ \bibinfo {pages} {253401} (\bibinfo {year} {2024}{\natexlab{a}})}\BibitemShut {NoStop}%
\bibitem [{\citenamefont {Halperin}(1983)}]{halperin83}%
  \BibitemOpen
  \bibfield  {author} {\bibinfo {author} {\bibfnamefont {B.~I.}\ \bibnamefont {Halperin}},\ }\bibfield  {title} {\bibinfo {title} {{Theory of the quantized Hall conductance}},\ }\href {https://doi.org/10.5169/seals-115362} {\bibfield  {journal} {\bibinfo  {journal} {Helv. Phys. Acta}\ }\textbf {\bibinfo {volume} {56}},\ \bibinfo {pages} {75} (\bibinfo {year} {1983})}\BibitemShut {NoStop}%
\bibitem [{\citenamefont {Kwan}\ \emph {et~al.}()\citenamefont {Kwan}, \citenamefont {Segura}, \citenamefont {Li}, \citenamefont {Blatz}, \citenamefont {Zhi}, \citenamefont {Bakkali-Hassani}, \citenamefont {Bohrdt}, \citenamefont {Greiter}, \citenamefont {Grusdt},\ and\ \citenamefont {Greiner}}]{kwan26}%
  \BibitemOpen
  \bibfield  {author} {\bibinfo {author} {\bibfnamefont {J.}~\bibnamefont {Kwan}}, \bibinfo {author} {\bibfnamefont {P.}~\bibnamefont {Segura}}, \bibinfo {author} {\bibfnamefont {Y.}~\bibnamefont {Li}}, \bibinfo {author} {\bibfnamefont {T.}~\bibnamefont {Blatz}}, \bibinfo {author} {\bibfnamefont {A.}~\bibnamefont {Zhi}}, \bibinfo {author} {\bibfnamefont {B.}~\bibnamefont {Bakkali-Hassani}}, \bibinfo {author} {\bibfnamefont {A.}~\bibnamefont {Bohrdt}}, \bibinfo {author} {\bibfnamefont {M.}~\bibnamefont {Greiter}}, \bibinfo {author} {\bibfnamefont {F.}~\bibnamefont {Grusdt}},\ and\ \bibinfo {author} {\bibfnamefont {M.}~\bibnamefont {Greiner}},\ }\href {https://arxiv.org/abs/2606.12409} {\bibinfo {title} {{A Pfaffian quantum Hall state of ultracold bosons}}},\ \Eprint {https://arxiv.org/abs/2606.12409} {arXiv:2606.12409} \BibitemShut {NoStop}%
\bibitem [{\citenamefont {Wu}\ \emph {et~al.}(2025)\citenamefont {Wu}, \citenamefont {Li}, \citenamefont {Goldman},\ and\ \citenamefont {Wang}}]{wu25}%
  \BibitemOpen
  \bibfield  {author} {\bibinfo {author} {\bibfnamefont {L.-N.}\ \bibnamefont {Wu}}, \bibinfo {author} {\bibfnamefont {X.}~\bibnamefont {Li}}, \bibinfo {author} {\bibfnamefont {N.}~\bibnamefont {Goldman}},\ and\ \bibinfo {author} {\bibfnamefont {B.}~\bibnamefont {Wang}},\ }\bibfield  {title} {\bibinfo {title} {{Optimal control for preparing fractional quantum Hall states in optical lattices}},\ }\href {https://doi.org/10.1103/PhysRevB.111.235111} {\bibfield  {journal} {\bibinfo  {journal} {Phys. Rev. B}\ }\textbf {\bibinfo {volume} {111}},\ \bibinfo {pages} {235111} (\bibinfo {year} {2025})}\BibitemShut {NoStop}%
\bibitem [{\citenamefont {Baur}\ \emph {et~al.}(2008)\citenamefont {Baur}, \citenamefont {Hazzard},\ and\ \citenamefont {Mueller}}]{baur08}%
  \BibitemOpen
  \bibfield  {author} {\bibinfo {author} {\bibfnamefont {S.~K.}\ \bibnamefont {Baur}}, \bibinfo {author} {\bibfnamefont {K.~R.~A.}\ \bibnamefont {Hazzard}},\ and\ \bibinfo {author} {\bibfnamefont {E.~J.}\ \bibnamefont {Mueller}},\ }\bibfield  {title} {\bibinfo {title} {{Stirring trapped atoms into fractional quantum Hall puddles}},\ }\href {https://doi.org/10.1103/PhysRevA.78.061608} {\bibfield  {journal} {\bibinfo  {journal} {Phys. Rev. A}\ }\textbf {\bibinfo {volume} {78}},\ \bibinfo {pages} {061608} (\bibinfo {year} {2008})}\BibitemShut {NoStop}%
\bibitem [{\citenamefont {Palm}\ \emph {et~al.}(2020)\citenamefont {Palm}, \citenamefont {Grusdt},\ and\ \citenamefont {Preiss}}]{palm20}%
  \BibitemOpen
  \bibfield  {author} {\bibinfo {author} {\bibfnamefont {L.}~\bibnamefont {Palm}}, \bibinfo {author} {\bibfnamefont {F.}~\bibnamefont {Grusdt}},\ and\ \bibinfo {author} {\bibfnamefont {P.~M.}\ \bibnamefont {Preiss}},\ }\bibfield  {title} {\bibinfo {title} {{Skyrmion ground states of rapidly rotating few-fermion systems}},\ }\href {https://doi.org/10.1088/1367-2630/aba30e} {\bibfield  {journal} {\bibinfo  {journal} {New J. Phys.}\ }\textbf {\bibinfo {volume} {22}},\ \bibinfo {pages} {083037} (\bibinfo {year} {2020})}\BibitemShut {NoStop}%
\bibitem [{\citenamefont {Popp}\ \emph {et~al.}(2004)\citenamefont {Popp}, \citenamefont {Paredes},\ and\ \citenamefont {Cirac}}]{popp04}%
  \BibitemOpen
  \bibfield  {author} {\bibinfo {author} {\bibfnamefont {M.}~\bibnamefont {Popp}}, \bibinfo {author} {\bibfnamefont {B.}~\bibnamefont {Paredes}},\ and\ \bibinfo {author} {\bibfnamefont {J.~I.}\ \bibnamefont {Cirac}},\ }\bibfield  {title} {\bibinfo {title} {{Adiabatic path to fractional quantum Hall states of a few bosonic atoms}},\ }\href {https://doi.org/10.1103/PhysRevA.70.053612} {\bibfield  {journal} {\bibinfo  {journal} {Phys. Rev. A}\ }\textbf {\bibinfo {volume} {70}},\ \bibinfo {pages} {053612} (\bibinfo {year} {2004})}\BibitemShut {NoStop}%
\bibitem [{\citenamefont {Andrade}\ \emph {et~al.}(2021)\citenamefont {Andrade}, \citenamefont {Kasper}, \citenamefont {Lewenstein}, \citenamefont {Weitenberg},\ and\ \citenamefont {Gra\ss{}}}]{andrade21}%
  \BibitemOpen
  \bibfield  {author} {\bibinfo {author} {\bibfnamefont {B.}~\bibnamefont {Andrade}}, \bibinfo {author} {\bibfnamefont {V.}~\bibnamefont {Kasper}}, \bibinfo {author} {\bibfnamefont {M.}~\bibnamefont {Lewenstein}}, \bibinfo {author} {\bibfnamefont {C.}~\bibnamefont {Weitenberg}},\ and\ \bibinfo {author} {\bibfnamefont {T.}~\bibnamefont {Gra\ss{}}},\ }\bibfield  {title} {\bibinfo {title} {{Preparation of the 1/2 Laughlin state with atoms in a rotating trap}},\ }\href {https://doi.org/10.1103/PhysRevA.103.063325} {\bibfield  {journal} {\bibinfo  {journal} {Phys. Rev. A}\ }\textbf {\bibinfo {volume} {103}},\ \bibinfo {pages} {063325} (\bibinfo {year} {2021})}\BibitemShut {NoStop}%
\bibitem [{\citenamefont {Wilkin}\ \emph {et~al.}(1998)\citenamefont {Wilkin}, \citenamefont {Gunn},\ and\ \citenamefont {Smith}}]{wilkin98}%
  \BibitemOpen
  \bibfield  {author} {\bibinfo {author} {\bibfnamefont {N.~K.}\ \bibnamefont {Wilkin}}, \bibinfo {author} {\bibfnamefont {J.~M.~F.}\ \bibnamefont {Gunn}},\ and\ \bibinfo {author} {\bibfnamefont {R.~A.}\ \bibnamefont {Smith}},\ }\bibfield  {title} {\bibinfo {title} {{Do Attractive Bosons Condense?}},\ }\href {https://doi.org/10.1103/PhysRevLett.80.2265} {\bibfield  {journal} {\bibinfo  {journal} {Phys. Rev. Lett.}\ }\textbf {\bibinfo {volume} {80}},\ \bibinfo {pages} {2265} (\bibinfo {year} {1998})}\BibitemShut {NoStop}%
\bibitem [{\citenamefont {Bayha}\ \emph {et~al.}(2020)\citenamefont {Bayha}, \citenamefont {Holten}, \citenamefont {Klemt}, \citenamefont {Subramanian}, \citenamefont {Bjerlin}, \citenamefont {Reimann}, \citenamefont {Bruun}, \citenamefont {Preiss},\ and\ \citenamefont {Jochim}}]{bayha20}%
  \BibitemOpen
  \bibfield  {author} {\bibinfo {author} {\bibfnamefont {L.}~\bibnamefont {Bayha}}, \bibinfo {author} {\bibfnamefont {M.}~\bibnamefont {Holten}}, \bibinfo {author} {\bibfnamefont {R.}~\bibnamefont {Klemt}}, \bibinfo {author} {\bibfnamefont {K.}~\bibnamefont {Subramanian}}, \bibinfo {author} {\bibfnamefont {J.}~\bibnamefont {Bjerlin}}, \bibinfo {author} {\bibfnamefont {S.~M.}\ \bibnamefont {Reimann}}, \bibinfo {author} {\bibfnamefont {G.~M.}\ \bibnamefont {Bruun}}, \bibinfo {author} {\bibfnamefont {P.~M.}\ \bibnamefont {Preiss}},\ and\ \bibinfo {author} {\bibfnamefont {S.}~\bibnamefont {Jochim}},\ }\bibfield  {title} {\bibinfo {title} {{Observing the emergence of a quantum phase transition shell by shell}},\ }\href {https://doi.org/10.1038/s41586-020-2936-y} {\bibfield  {journal} {\bibinfo  {journal} {Nature (London)}\ }\textbf {\bibinfo {volume} {587}},\ \bibinfo {pages} {583} (\bibinfo {year} {2020})}\BibitemShut {NoStop}%
\bibitem [{\citenamefont {Yannouleas}\ and\ \citenamefont {Landman}(2020)}]{yannouleas20}%
  \BibitemOpen
  \bibfield  {author} {\bibinfo {author} {\bibfnamefont {C.}~\bibnamefont {Yannouleas}}\ and\ \bibinfo {author} {\bibfnamefont {U.}~\bibnamefont {Landman}},\ }\bibfield  {title} {\bibinfo {title} {{Fractional quantum Hall physics and higher-order momentum correlations in a few spinful fermionic contact-interacting ultracold atoms in rotating traps}},\ }\href {https://doi.org/10.1103/PhysRevA.102.043317} {\bibfield  {journal} {\bibinfo  {journal} {Phys. Rev. A}\ }\textbf {\bibinfo {volume} {102}},\ \bibinfo {pages} {043317} (\bibinfo {year} {2020})}\BibitemShut {NoStop}%
\bibitem [{\citenamefont {Hofmann}\ and\ \citenamefont {Zwerger}(2023)}]{hofmann23}%
  \BibitemOpen
  \bibfield  {author} {\bibinfo {author} {\bibfnamefont {J.}~\bibnamefont {Hofmann}}\ and\ \bibinfo {author} {\bibfnamefont {W.}~\bibnamefont {Zwerger}},\ }\bibfield  {title} {\bibinfo {title} {Scale {Invariance} in the {Lowest} {Landau} {Level}},\ }\href {https://doi.org/10.5802/crphys.137} {\bibfield  {journal} {\bibinfo  {journal} {C. R. Phys.}\ }\textbf {\bibinfo {volume} {24}},\ \bibinfo {pages} {69} (\bibinfo {year} {2023})}\BibitemShut {NoStop}%
\bibitem [{\citenamefont {Regnault}\ and\ \citenamefont {Jolicoeur}(2004)}]{Regnault04}%
  \BibitemOpen
  \bibfield  {author} {\bibinfo {author} {\bibfnamefont {N.}~\bibnamefont {Regnault}}\ and\ \bibinfo {author} {\bibfnamefont {T.}~\bibnamefont {Jolicoeur}},\ }\bibfield  {title} {\bibinfo {title} {{Quantum Hall fractions in ultracold fermionic vapors}},\ }\href {https://doi.org/10.1103/PhysRevB.70.241307} {\bibfield  {journal} {\bibinfo  {journal} {Phys. Rev. B}\ }\textbf {\bibinfo {volume} {70}},\ \bibinfo {pages} {241307} (\bibinfo {year} {2004})}\BibitemShut {NoStop}%
\bibitem [{\citenamefont {Sakurai}(1994)}]{sakurai94}%
  \BibitemOpen
  \bibfield  {author} {\bibinfo {author} {\bibfnamefont {J.~J.}\ \bibnamefont {Sakurai}},\ }\href {https://doi.org/10.1017/9781108587280} {\emph {\bibinfo {title} {{Modern Quantum Mechanics}}}}\ (\bibinfo  {publisher} {Addison-Wesley, Reading, Massachusetts},\ \bibinfo {year} {1994})\BibitemShut {NoStop}%
\bibitem [{\citenamefont {Lunt}\ \emph {et~al.}(2024{\natexlab{b}})\citenamefont {Lunt}, \citenamefont {Hill}, \citenamefont {Reiter}, \citenamefont {Preiss}, \citenamefont {Ga\l{}ka},\ and\ \citenamefont {Jochim}}]{lunt24b}%
  \BibitemOpen
  \bibfield  {author} {\bibinfo {author} {\bibfnamefont {P.}~\bibnamefont {Lunt}}, \bibinfo {author} {\bibfnamefont {P.}~\bibnamefont {Hill}}, \bibinfo {author} {\bibfnamefont {J.}~\bibnamefont {Reiter}}, \bibinfo {author} {\bibfnamefont {P.~M.}\ \bibnamefont {Preiss}}, \bibinfo {author} {\bibfnamefont {M.}~\bibnamefont {Ga\l{}ka}},\ and\ \bibinfo {author} {\bibfnamefont {S.}~\bibnamefont {Jochim}},\ }\bibfield  {title} {\bibinfo {title} {{Engineering single-atom angular momentum eigenstates in an optical tweezer}},\ }\href {https://doi.org/10.1103/PhysRevA.110.063315} {\bibfield  {journal} {\bibinfo  {journal} {Phys. Rev. A}\ }\textbf {\bibinfo {volume} {110}},\ \bibinfo {pages} {063315} (\bibinfo {year} {2024}{\natexlab{b}})}\BibitemShut {NoStop}%
\bibitem [{Note1()}]{Note1}%
  \BibitemOpen
  \bibinfo {note} {The Halperin wave function is given by \protect \mbox {$\Psi _{(m_1,m_2,m_3)}(z_{1\uparrow }, \protect \ldots , z_{1\downarrow },\protect \ldots ) =\DOTSB \prod@ \slimits@ _{i<j=1}^{N_{\uparrow }} (z_{i\uparrow } - z_{j\uparrow })^{m_1}$} $ \DOTSB \prod@ \slimits@ _{i<j=1}^{N_{\downarrow }} (z_{i\downarrow } -z_{j\downarrow })^{m_2}\DOTSB \prod@ \slimits@ _{i,j=1}^{N_{\uparrow }, N_{\downarrow }} (z_{i\uparrow } - z_{j\downarrow })^{m_3}$, and the two-particle state discussed in the text corresponds to \protect \mbox {$N_{\uparrow }=N_{\downarrow }=1$} and \protect \mbox {$(m_1,m_2,m_3) = (0,0,2)$}.}\BibitemShut {Stop}%
\bibitem [{\citenamefont {Bergschneider}\ \emph {et~al.}(2018)\citenamefont {Bergschneider}, \citenamefont {Klinkhamer}, \citenamefont {Becher}, \citenamefont {Klemt}, \citenamefont {Z\"urn}, \citenamefont {Preiss},\ and\ \citenamefont {Jochim}}]{bergschneider18}%
  \BibitemOpen
  \bibfield  {author} {\bibinfo {author} {\bibfnamefont {A.}~\bibnamefont {Bergschneider}}, \bibinfo {author} {\bibfnamefont {V.~M.}\ \bibnamefont {Klinkhamer}}, \bibinfo {author} {\bibfnamefont {J.~H.}\ \bibnamefont {Becher}}, \bibinfo {author} {\bibfnamefont {R.}~\bibnamefont {Klemt}}, \bibinfo {author} {\bibfnamefont {G.}~\bibnamefont {Z\"urn}}, \bibinfo {author} {\bibfnamefont {P.~M.}\ \bibnamefont {Preiss}},\ and\ \bibinfo {author} {\bibfnamefont {S.}~\bibnamefont {Jochim}},\ }\bibfield  {title} {\bibinfo {title} {{Spin-resolved single-atom imaging of $^{6}\mathrm{Li}$ in free space}},\ }\href {https://doi.org/10.1103/PhysRevA.97.063613} {\bibfield  {journal} {\bibinfo  {journal} {Phys. Rev. A}\ }\textbf {\bibinfo {volume} {97}},\ \bibinfo {pages} {063613} (\bibinfo {year} {2018})}\BibitemShut {NoStop}%
\bibitem [{\citenamefont {Holten}\ \emph {et~al.}(2021)\citenamefont {Holten}, \citenamefont {Bayha}, \citenamefont {Subramanian}, \citenamefont {Heintze}, \citenamefont {Preiss},\ and\ \citenamefont {Jochim}}]{holten21a}%
  \BibitemOpen
  \bibfield  {author} {\bibinfo {author} {\bibfnamefont {M.}~\bibnamefont {Holten}}, \bibinfo {author} {\bibfnamefont {L.}~\bibnamefont {Bayha}}, \bibinfo {author} {\bibfnamefont {K.}~\bibnamefont {Subramanian}}, \bibinfo {author} {\bibfnamefont {C.}~\bibnamefont {Heintze}}, \bibinfo {author} {\bibfnamefont {P.~M.}\ \bibnamefont {Preiss}},\ and\ \bibinfo {author} {\bibfnamefont {S.}~\bibnamefont {Jochim}},\ }\bibfield  {title} {\bibinfo {title} {{Observation of Pauli Crystals}},\ }\href {https://doi.org/10.1103/PhysRevLett.126.020401} {\bibfield  {journal} {\bibinfo  {journal} {Phys. Rev. Lett.}\ }\textbf {\bibinfo {volume} {126}},\ \bibinfo {pages} {020401} (\bibinfo {year} {2021})}\BibitemShut {NoStop}%
\bibitem [{\citenamefont {Holten}\ \emph {et~al.}(2022)\citenamefont {Holten}, \citenamefont {Bayha}, \citenamefont {Subramanian}, \citenamefont {Brandstetter}, \citenamefont {Heintze}, \citenamefont {Lunt}, \citenamefont {Preiss},\ and\ \citenamefont {Jochim}}]{holten21b}%
  \BibitemOpen
  \bibfield  {author} {\bibinfo {author} {\bibfnamefont {M.}~\bibnamefont {Holten}}, \bibinfo {author} {\bibfnamefont {L.}~\bibnamefont {Bayha}}, \bibinfo {author} {\bibfnamefont {K.}~\bibnamefont {Subramanian}}, \bibinfo {author} {\bibfnamefont {S.}~\bibnamefont {Brandstetter}}, \bibinfo {author} {\bibfnamefont {C.}~\bibnamefont {Heintze}}, \bibinfo {author} {\bibfnamefont {P.}~\bibnamefont {Lunt}}, \bibinfo {author} {\bibfnamefont {P.~M.}\ \bibnamefont {Preiss}},\ and\ \bibinfo {author} {\bibfnamefont {S.}~\bibnamefont {Jochim}},\ }\bibfield  {title} {\bibinfo {title} {{Observation of Cooper pairs in a mesoscopic two-dimensional Fermi gas}},\ }\href {https://doi.org/10.1038/s41586-022-04678-1} {\bibfield  {journal} {\bibinfo  {journal} {Nature (London)}\ }\textbf {\bibinfo {volume} {606}},\ \bibinfo {pages} {287} (\bibinfo {year} {2022})}\BibitemShut {NoStop}%
\bibitem [{Note2()}]{Note2}%
  \BibitemOpen
  \bibinfo {note} {Explicitly, the single-particle wave function of a state \protect \mbox {$j=\{n_j\geq 0,k_j \geq 0\}$} is $ \ell _{ho} \phi _j(z,\protect \bar {z}) = \protect \sqrt {\protect \frac {\min (n_j,k_j)!}{\pi \max (n_j,k_j)!}} \ (z/\ell _{ho})^{|k_j-n_j|} e^{-\protect \bar {z}z/2\ell _{ho}^2} \protect \, L_{\min (n_j,k_j)}^{|k_j-n_j|}(\protect \bar {z}z/\ell _{ho}^2)$, for $n_j\leq k_j $ and else $z$ exchanged with $\protect \bar {z}$, where $\ell _{ho}$ is the harmonic oscillator length.}\BibitemShut {Stop}%
\bibitem [{\citenamefont {Zwerger}(2016)}]{zwerger16}%
  \BibitemOpen
  \bibfield  {author} {\bibinfo {author} {\bibfnamefont {W.}~\bibnamefont {Zwerger}},\ }\bibfield  {title} {\bibinfo {title} {{Strongly interacting {Fermi} gases}},\ }in\ \href {https://doi.org/10.3254/978-1-61499-694-1-63} {\emph {\bibinfo {booktitle} {Proceedings of the International School of Physics ``Enrico Fermi -- Course 191 "Quantum Matter at Ultralow Temperatures"}}},\ \bibinfo {editor} {edited by\ \bibinfo {editor} {\bibfnamefont {S.}~\bibnamefont {Stringari}}, \bibinfo {editor} {\bibfnamefont {M.}~\bibnamefont {Inguscio}}, \bibinfo {editor} {\bibfnamefont {W.}~\bibnamefont {Ketterle}},\ and\ \bibinfo {editor} {\bibfnamefont {G.}~\bibnamefont {Roati}}}\ (\bibinfo  {publisher} {IOS Press, Amsterdam; SIF Bologna},\ \bibinfo {year} {2016})\ pp.\ \bibinfo {pages} {63--142}\BibitemShut {NoStop}%
\bibitem [{\citenamefont {Hofmann}\ and\ \citenamefont {Zwerger}(2021)}]{hofmann21}%
  \BibitemOpen
  \bibfield  {author} {\bibinfo {author} {\bibfnamefont {J.}~\bibnamefont {Hofmann}}\ and\ \bibinfo {author} {\bibfnamefont {W.}~\bibnamefont {Zwerger}},\ }\bibfield  {title} {\bibinfo {title} {{Universal relations for dipolar quantum gases}},\ }\href {https://doi.org/10.1103/PhysRevResearch.3.013088} {\bibfield  {journal} {\bibinfo  {journal} {Phys. Rev. Res.}\ }\textbf {\bibinfo {volume} {3}},\ \bibinfo {pages} {013088} (\bibinfo {year} {2021})}\BibitemShut {NoStop}%
\bibitem [{\citenamefont {Haldane}(1983)}]{Haldane83}%
  \BibitemOpen
  \bibfield  {author} {\bibinfo {author} {\bibfnamefont {F.~D.~M.}\ \bibnamefont {Haldane}},\ }\bibfield  {title} {\bibinfo {title} {{Fractional Quantization of the Hall Effect: A Hierarchy of Incompressible Quantum Fluid States}},\ }\href {https://doi.org/10.1103/PhysRevLett.51.605} {\bibfield  {journal} {\bibinfo  {journal} {Phys. Rev. Lett.}\ }\textbf {\bibinfo {volume} {51}},\ \bibinfo {pages} {605} (\bibinfo {year} {1983})}\BibitemShut {NoStop}%
\bibitem [{\citenamefont {Craps}\ \emph {et~al.}(2024)\citenamefont {Craps}, \citenamefont {De~Clerck}, \citenamefont {Evnin},\ and\ \citenamefont {Pavlov}}]{craps24}%
  \BibitemOpen
  \bibfield  {author} {\bibinfo {author} {\bibfnamefont {B.}~\bibnamefont {Craps}}, \bibinfo {author} {\bibfnamefont {M.}~\bibnamefont {De~Clerck}}, \bibinfo {author} {\bibfnamefont {O.}~\bibnamefont {Evnin}},\ and\ \bibinfo {author} {\bibfnamefont {M.}~\bibnamefont {Pavlov}},\ }\bibfield  {title} {\bibinfo {title} {{Phase-space localization at the lowest Landau level}},\ }\href {https://doi.org/10.1103/PhysRevA.110.063325} {\bibfield  {journal} {\bibinfo  {journal} {Phys. Rev. A}\ }\textbf {\bibinfo {volume} {110}},\ \bibinfo {pages} {063325} (\bibinfo {year} {2024})}\BibitemShut {NoStop}%
\bibitem [{foo()}]{footnote}%
  \BibitemOpen
  \href@noop {} {}\bibinfo {note} {{See End Matter for details.}}\BibitemShut {Stop}%
\bibitem [{\citenamefont {Trugman}\ and\ \citenamefont {Kivelson}(1985)}]{trugman85}%
  \BibitemOpen
  \bibfield  {author} {\bibinfo {author} {\bibfnamefont {S.~A.}\ \bibnamefont {Trugman}}\ and\ \bibinfo {author} {\bibfnamefont {S.}~\bibnamefont {Kivelson}},\ }\bibfield  {title} {\bibinfo {title} {{Exact results for the fractional quantum Hall effect with general interactions}},\ }\href {https://doi.org/10.1103/PhysRevB.31.5280} {\bibfield  {journal} {\bibinfo  {journal} {Phys. Rev. B}\ }\textbf {\bibinfo {volume} {31}},\ \bibinfo {pages} {5280} (\bibinfo {year} {1985})}\BibitemShut {NoStop}%
\bibitem [{\citenamefont {Xiang}\ \emph {et~al.}(2025)\citenamefont {Xiang}, \citenamefont {Cruz-Col\'on}, \citenamefont {Chua}, \citenamefont {Milner}, \citenamefont {de~Hond}, \citenamefont {Fricke},\ and\ \citenamefont {Ketterle}}]{Xiang25}%
  \BibitemOpen
  \bibfield  {author} {\bibinfo {author} {\bibfnamefont {J.}~\bibnamefont {Xiang}}, \bibinfo {author} {\bibfnamefont {E.}~\bibnamefont {Cruz-Col\'on}}, \bibinfo {author} {\bibfnamefont {C.~C.}\ \bibnamefont {Chua}}, \bibinfo {author} {\bibfnamefont {W.~R.}\ \bibnamefont {Milner}}, \bibinfo {author} {\bibfnamefont {J.}~\bibnamefont {de~Hond}}, \bibinfo {author} {\bibfnamefont {J.~F.}\ \bibnamefont {Fricke}},\ and\ \bibinfo {author} {\bibfnamefont {W.}~\bibnamefont {Ketterle}},\ }\bibfield  {title} {\bibinfo {title} {{In Situ Imaging of the Thermal de Broglie Wavelength in an Ultracold Bose Gas}},\ }\href {https://doi.org/10.1103/PhysRevLett.134.183401} {\bibfield  {journal} {\bibinfo  {journal} {Phys. Rev. Lett.}\ }\textbf {\bibinfo {volume} {134}},\ \bibinfo {pages} {183401} (\bibinfo {year} {2025})}\BibitemShut {NoStop}%
\bibitem [{\citenamefont {Yao}\ \emph {et~al.}(2025)\citenamefont {Yao}, \citenamefont {Chi}, \citenamefont {Wang}, \citenamefont {Fletcher},\ and\ \citenamefont {Zwierlein}}]{Yao25}%
  \BibitemOpen
  \bibfield  {author} {\bibinfo {author} {\bibfnamefont {R.}~\bibnamefont {Yao}}, \bibinfo {author} {\bibfnamefont {S.}~\bibnamefont {Chi}}, \bibinfo {author} {\bibfnamefont {M.}~\bibnamefont {Wang}}, \bibinfo {author} {\bibfnamefont {R.~J.}\ \bibnamefont {Fletcher}},\ and\ \bibinfo {author} {\bibfnamefont {M.}~\bibnamefont {Zwierlein}},\ }\bibfield  {title} {\bibinfo {title} {{Measuring Pair Correlations in Bose and Fermi Gases via Atom-Resolved Microscopy}},\ }\href {https://doi.org/10.1103/PhysRevLett.134.183402} {\bibfield  {journal} {\bibinfo  {journal} {Phys. Rev. Lett.}\ }\textbf {\bibinfo {volume} {134}},\ \bibinfo {pages} {183402} (\bibinfo {year} {2025})}\BibitemShut {NoStop}%
\bibitem [{\citenamefont {de~Jongh}\ \emph {et~al.}(2025)\citenamefont {de~Jongh}, \citenamefont {Verstraten}, \citenamefont {Dixmerias}, \citenamefont {Daix}, \citenamefont {Peaudecerf},\ and\ \citenamefont {Yefsah}}]{deJongh25}%
  \BibitemOpen
  \bibfield  {author} {\bibinfo {author} {\bibfnamefont {T.}~\bibnamefont {de~Jongh}}, \bibinfo {author} {\bibfnamefont {J.}~\bibnamefont {Verstraten}}, \bibinfo {author} {\bibfnamefont {M.}~\bibnamefont {Dixmerias}}, \bibinfo {author} {\bibfnamefont {C.}~\bibnamefont {Daix}}, \bibinfo {author} {\bibfnamefont {B.}~\bibnamefont {Peaudecerf}},\ and\ \bibinfo {author} {\bibfnamefont {T.}~\bibnamefont {Yefsah}},\ }\bibfield  {title} {\bibinfo {title} {{Quantum Gas Microscopy of Fermions in the Continuum}},\ }\href {https://doi.org/10.1103/PhysRevLett.134.183403} {\bibfield  {journal} {\bibinfo  {journal} {Phys. Rev. Lett.}\ }\textbf {\bibinfo {volume} {134}},\ \bibinfo {pages} {183403} (\bibinfo {year} {2025})}\BibitemShut {NoStop}%
\bibitem [{\citenamefont {Bekassy}\ and\ \citenamefont {Hofmann}(2022)}]{bekassy22}%
  \BibitemOpen
  \bibfield  {author} {\bibinfo {author} {\bibfnamefont {V.}~\bibnamefont {Bekassy}}\ and\ \bibinfo {author} {\bibfnamefont {J.}~\bibnamefont {Hofmann}},\ }\bibfield  {title} {\bibinfo {title} {{Nonrelativistic Conformal Invariance in Mesoscopic Two-Dimensional Fermi Gases}},\ }\href {https://doi.org/10.1103/PhysRevLett.128.193401} {\bibfield  {journal} {\bibinfo  {journal} {Phys. Rev. Lett.}\ }\textbf {\bibinfo {volume} {128}},\ \bibinfo {pages} {193401} (\bibinfo {year} {2022})}\BibitemShut {NoStop}%
\bibitem [{\citenamefont {Bekassy}\ and\ \citenamefont {Hofmann}(2024)}]{bekassy24}%
  \BibitemOpen
  \bibfield  {author} {\bibinfo {author} {\bibfnamefont {V.}~\bibnamefont {Bekassy}}\ and\ \bibinfo {author} {\bibfnamefont {J.}~\bibnamefont {Hofmann}},\ }\bibfield  {title} {\bibinfo {title} {{Scale and conformal invariance in rotating interacting few-fermion systems}},\ }\href {https://doi.org/10.1103/PhysRevResearch.6.023279} {\bibfield  {journal} {\bibinfo  {journal} {Phys. Rev. Res.}\ }\textbf {\bibinfo {volume} {6}},\ \bibinfo {pages} {023279} (\bibinfo {year} {2024})}\BibitemShut {NoStop}%
\bibitem [{\citenamefont {Girvin}\ \emph {et~al.}(1986)\citenamefont {Girvin}, \citenamefont {MacDonald},\ and\ \citenamefont {Platzman}}]{girvin86}%
  \BibitemOpen
  \bibfield  {author} {\bibinfo {author} {\bibfnamefont {S.~M.}\ \bibnamefont {Girvin}}, \bibinfo {author} {\bibfnamefont {A.~H.}\ \bibnamefont {MacDonald}},\ and\ \bibinfo {author} {\bibfnamefont {P.~M.}\ \bibnamefont {Platzman}},\ }\bibfield  {title} {\bibinfo {title} {{Magneto-roton theory of collective excitations in the fractional quantum Hall effect}},\ }\href {https://doi.org/10.1103/PhysRevB.33.2481} {\bibfield  {journal} {\bibinfo  {journal} {Phys. Rev. B}\ }\textbf {\bibinfo {volume} {33}},\ \bibinfo {pages} {2481} (\bibinfo {year} {1986})}\BibitemShut {NoStop}%
\bibitem [{\citenamefont {Kukushkin}\ \emph {et~al.}(2009)\citenamefont {Kukushkin}, \citenamefont {Smet}, \citenamefont {Scarola}, \citenamefont {Umansky},\ and\ \citenamefont {von Klitzing}}]{kukushkin09}%
  \BibitemOpen
  \bibfield  {author} {\bibinfo {author} {\bibfnamefont {I.~V.}\ \bibnamefont {Kukushkin}}, \bibinfo {author} {\bibfnamefont {J.~H.}\ \bibnamefont {Smet}}, \bibinfo {author} {\bibfnamefont {V.~W.}\ \bibnamefont {Scarola}}, \bibinfo {author} {\bibfnamefont {V.}~\bibnamefont {Umansky}},\ and\ \bibinfo {author} {\bibfnamefont {K.}~\bibnamefont {von Klitzing}},\ }\bibfield  {title} {\bibinfo {title} {{Dispersion of the Excitations of Fractional Quantum Hall States}},\ }\href {https://doi.org/10.1126/science.1171472} {\bibfield  {journal} {\bibinfo  {journal} {Science}\ }\textbf {\bibinfo {volume} {324}},\ \bibinfo {pages} {1044} (\bibinfo {year} {2009})}\BibitemShut {NoStop}%
\bibitem [{\citenamefont {Haldane}(2011)}]{haldane11}%
  \BibitemOpen
  \bibfield  {author} {\bibinfo {author} {\bibfnamefont {F.~D.~M.}\ \bibnamefont {Haldane}},\ }\bibfield  {title} {\bibinfo {title} {{Geometrical Description of the Fractional Quantum Hall Effect}},\ }\href {https://doi.org/10.1103/PhysRevLett.107.116801} {\bibfield  {journal} {\bibinfo  {journal} {Phys. Rev. Lett.}\ }\textbf {\bibinfo {volume} {107}},\ \bibinfo {pages} {116801} (\bibinfo {year} {2011})}\BibitemShut {NoStop}%
\bibitem [{\citenamefont {Liou}\ \emph {et~al.}(2019)\citenamefont {Liou}, \citenamefont {Haldane}, \citenamefont {Yang},\ and\ \citenamefont {Rezayi}}]{liou19}%
  \BibitemOpen
  \bibfield  {author} {\bibinfo {author} {\bibfnamefont {S.-F.}\ \bibnamefont {Liou}}, \bibinfo {author} {\bibfnamefont {F.~D.~M.}\ \bibnamefont {Haldane}}, \bibinfo {author} {\bibfnamefont {K.}~\bibnamefont {Yang}},\ and\ \bibinfo {author} {\bibfnamefont {E.~H.}\ \bibnamefont {Rezayi}},\ }\bibfield  {title} {\bibinfo {title} {{Chiral Gravitons in Fractional Quantum Hall Liquids}},\ }\href {https://doi.org/10.1103/PhysRevLett.123.146801} {\bibfield  {journal} {\bibinfo  {journal} {Phys. Rev. Lett.}\ }\textbf {\bibinfo {volume} {123}},\ \bibinfo {pages} {146801} (\bibinfo {year} {2019})}\BibitemShut {NoStop}%
\bibitem [{\citenamefont {Pinczuk}\ \emph {et~al.}(1993)\citenamefont {Pinczuk}, \citenamefont {Dennis}, \citenamefont {Pfeiffer},\ and\ \citenamefont {West}}]{pinczuk93}%
  \BibitemOpen
  \bibfield  {author} {\bibinfo {author} {\bibfnamefont {A.}~\bibnamefont {Pinczuk}}, \bibinfo {author} {\bibfnamefont {B.~S.}\ \bibnamefont {Dennis}}, \bibinfo {author} {\bibfnamefont {L.~N.}\ \bibnamefont {Pfeiffer}},\ and\ \bibinfo {author} {\bibfnamefont {K.}~\bibnamefont {West}},\ }\bibfield  {title} {\bibinfo {title} {{Observation of collective excitations in the fractional quantum Hall effect}},\ }\href {https://doi.org/10.1103/PhysRevLett.70.3983} {\bibfield  {journal} {\bibinfo  {journal} {Phys. Rev. Lett.}\ }\textbf {\bibinfo {volume} {70}},\ \bibinfo {pages} {3983} (\bibinfo {year} {1993})}\BibitemShut {NoStop}%
\bibitem [{\citenamefont {Liang}\ \emph {et~al.}(2024)\citenamefont {Liang}, \citenamefont {Liu}, \citenamefont {Yang}, \citenamefont {Huang}, \citenamefont {Wurstbauer}, \citenamefont {Dean}, \citenamefont {West}, \citenamefont {Pfeiffer}, \citenamefont {Du},\ and\ \citenamefont {Pinczuk}}]{liang24}%
  \BibitemOpen
  \bibfield  {author} {\bibinfo {author} {\bibfnamefont {J.}~\bibnamefont {Liang}}, \bibinfo {author} {\bibfnamefont {Z.}~\bibnamefont {Liu}}, \bibinfo {author} {\bibfnamefont {Z.}~\bibnamefont {Yang}}, \bibinfo {author} {\bibfnamefont {Y.}~\bibnamefont {Huang}}, \bibinfo {author} {\bibfnamefont {U.}~\bibnamefont {Wurstbauer}}, \bibinfo {author} {\bibfnamefont {C.~R.}\ \bibnamefont {Dean}}, \bibinfo {author} {\bibfnamefont {K.~W.}\ \bibnamefont {West}}, \bibinfo {author} {\bibfnamefont {L.~N.}\ \bibnamefont {Pfeiffer}}, \bibinfo {author} {\bibfnamefont {L.}~\bibnamefont {Du}},\ and\ \bibinfo {author} {\bibfnamefont {A.}~\bibnamefont {Pinczuk}},\ }\bibfield  {title} {\bibinfo {title} {{Evidence for chiral graviton modes in fractional quantum {Hall} liquids}},\ }\href {https://doi.org/10.1038/s41586-024-07201-w} {\bibfield  {journal} {\bibinfo  {journal} {Nature (London)}\ }\textbf {\bibinfo {volume} {628}},\ \bibinfo {pages} {78} (\bibinfo {year} {2024})}\BibitemShut {NoStop}%
\bibitem [{\citenamefont {Wigner}(1934)}]{wigner34}%
  \BibitemOpen
  \bibfield  {author} {\bibinfo {author} {\bibfnamefont {E.}~\bibnamefont {Wigner}},\ }\bibfield  {title} {\bibinfo {title} {{On the Interaction of Electrons in Metals}},\ }\href {https://doi.org/10.1103/PhysRev.46.1002} {\bibfield  {journal} {\bibinfo  {journal} {Phys. Rev.}\ }\textbf {\bibinfo {volume} {46}},\ \bibinfo {pages} {1002} (\bibinfo {year} {1934})}\BibitemShut {NoStop}%
\bibitem [{\citenamefont {Saarikoski}\ \emph {et~al.}(2010)\citenamefont {Saarikoski}, \citenamefont {Reimann}, \citenamefont {Harju},\ and\ \citenamefont {Manninen}}]{Saarikoski10}%
  \BibitemOpen
  \bibfield  {author} {\bibinfo {author} {\bibfnamefont {H.}~\bibnamefont {Saarikoski}}, \bibinfo {author} {\bibfnamefont {S.~M.}\ \bibnamefont {Reimann}}, \bibinfo {author} {\bibfnamefont {A.}~\bibnamefont {Harju}},\ and\ \bibinfo {author} {\bibfnamefont {M.}~\bibnamefont {Manninen}},\ }\bibfield  {title} {\bibinfo {title} {{Vortices in quantum droplets: Analogies between boson and fermion systems}},\ }\href {https://doi.org/10.1103/RevModPhys.82.2785} {\bibfield  {journal} {\bibinfo  {journal} {Rev. Mod. Phys.}\ }\textbf {\bibinfo {volume} {82}},\ \bibinfo {pages} {2785} (\bibinfo {year} {2010})}\BibitemShut {NoStop}%
\bibitem [{\citenamefont {Reimann}\ and\ \citenamefont {Manninen}(2002)}]{reimann02}%
  \BibitemOpen
  \bibfield  {author} {\bibinfo {author} {\bibfnamefont {S.~M.}\ \bibnamefont {Reimann}}\ and\ \bibinfo {author} {\bibfnamefont {M.}~\bibnamefont {Manninen}},\ }\bibfield  {title} {\bibinfo {title} {{Electronic structure of quantum dots}},\ }\href {https://doi.org/10.1103/RevModPhys.74.1283} {\bibfield  {journal} {\bibinfo  {journal} {Rev. Mod. Phys.}\ }\textbf {\bibinfo {volume} {74}},\ \bibinfo {pages} {1283} (\bibinfo {year} {2002})}\BibitemShut {NoStop}%
\bibitem [{\citenamefont {Leinaas}\ and\ \citenamefont {Myrheim}(1977)}]{Leinaas77}%
  \BibitemOpen
  \bibfield  {author} {\bibinfo {author} {\bibfnamefont {J.}~\bibnamefont {Leinaas}}\ and\ \bibinfo {author} {\bibfnamefont {J.}~\bibnamefont {Myrheim}},\ }\bibfield  {title} {\bibinfo {title} {{On the theory of identical particles}},\ }\href {https://doi.org/10.1007/BF02727953} {\bibfield  {journal} {\bibinfo  {journal} {Il Nuovo Cimento B}\ }\textbf {\bibinfo {volume} {37}},\ \bibinfo {pages} {1} (\bibinfo {year} {1977})}\BibitemShut {NoStop}%
\bibitem [{\citenamefont {Goldin}\ \emph {et~al.}(1980)\citenamefont {Goldin}, \citenamefont {Menikoff},\ and\ \citenamefont {Sharp}}]{goldin80}%
  \BibitemOpen
  \bibfield  {author} {\bibinfo {author} {\bibfnamefont {G.~A.}\ \bibnamefont {Goldin}}, \bibinfo {author} {\bibfnamefont {R.}~\bibnamefont {Menikoff}},\ and\ \bibinfo {author} {\bibfnamefont {D.~H.}\ \bibnamefont {Sharp}},\ }\bibfield  {title} {\bibinfo {title} {{Particle statistics from induced representations of a local current group}},\ }\href {https://doi.org/10.1063/1.524510} {\bibfield  {journal} {\bibinfo  {journal} {J. Math. Phys. (N.Y.)}\ }\textbf {\bibinfo {volume} {21}},\ \bibinfo {pages} {650} (\bibinfo {year} {1980})}\BibitemShut {NoStop}%
\bibitem [{\citenamefont {Wilczek}(1982)}]{wilczek82}%
  \BibitemOpen
  \bibfield  {author} {\bibinfo {author} {\bibfnamefont {F.}~\bibnamefont {Wilczek}},\ }\bibfield  {title} {\bibinfo {title} {{Quantum Mechanics of Fractional-Spin Particles}},\ }\href {https://doi.org/10.1103/PhysRevLett.49.957} {\bibfield  {journal} {\bibinfo  {journal} {Phys. Rev. Lett.}\ }\textbf {\bibinfo {volume} {49}},\ \bibinfo {pages} {957} (\bibinfo {year} {1982})}\BibitemShut {NoStop}%
\bibitem [{\citenamefont {Arovas}\ \emph {et~al.}(1984)\citenamefont {Arovas}, \citenamefont {Schrieffer},\ and\ \citenamefont {Wilczek}}]{arovas84}%
  \BibitemOpen
  \bibfield  {author} {\bibinfo {author} {\bibfnamefont {D.}~\bibnamefont {Arovas}}, \bibinfo {author} {\bibfnamefont {J.~R.}\ \bibnamefont {Schrieffer}},\ and\ \bibinfo {author} {\bibfnamefont {F.}~\bibnamefont {Wilczek}},\ }\bibfield  {title} {\bibinfo {title} {{Fractional Statistics and the Quantum Hall Effect}},\ }\href {https://doi.org/10.1103/PhysRevLett.53.722} {\bibfield  {journal} {\bibinfo  {journal} {Phys. Rev. Lett.}\ }\textbf {\bibinfo {volume} {53}},\ \bibinfo {pages} {722} (\bibinfo {year} {1984})}\BibitemShut {NoStop}%
\bibitem [{\citenamefont {Cooper}\ \emph {et~al.}(2005)\citenamefont {Cooper}, \citenamefont {Rezayi},\ and\ \citenamefont {Simon}}]{cooper05}%
  \BibitemOpen
  \bibfield  {author} {\bibinfo {author} {\bibfnamefont {N.~R.}\ \bibnamefont {Cooper}}, \bibinfo {author} {\bibfnamefont {E.~H.}\ \bibnamefont {Rezayi}},\ and\ \bibinfo {author} {\bibfnamefont {S.~H.}\ \bibnamefont {Simon}},\ }\bibfield  {title} {\bibinfo {title} {{Vortex Lattices in Rotating Atomic Bose Gases with Dipolar Interactions}},\ }\href {https://doi.org/10.1103/PhysRevLett.95.200402} {\bibfield  {journal} {\bibinfo  {journal} {Phys. Rev. Lett.}\ }\textbf {\bibinfo {volume} {95}},\ \bibinfo {pages} {200402} (\bibinfo {year} {2005})}\BibitemShut {NoStop}%
\bibitem [{\citenamefont {Haldane}(1987)}]{duncan87}%
  \BibitemOpen
  \bibfield  {author} {\bibinfo {author} {\bibfnamefont {F.~D.~M.}\ \bibnamefont {Haldane}},\ }\bibinfo {title} {{The Hierarchy of Fractional States and Numerical Studies}},\ in\ \href {https://doi.org/10.1007/978-1-4684-0499-9_8} {\emph {\bibinfo {booktitle} {The Quantum Hall Effect}}},\ \bibinfo {editor} {edited by\ \bibinfo {editor} {\bibfnamefont {R.~E.}\ \bibnamefont {Prange}}\ and\ \bibinfo {editor} {\bibfnamefont {S.~M.}\ \bibnamefont {Girvin}}}\ (\bibinfo  {publisher} {Springer US},\ \bibinfo {address} {New York, NY},\ \bibinfo {year} {1987})\ pp.\ \bibinfo {pages} {303--352}\BibitemShut {NoStop}%
\bibitem [{\citenamefont {Girvin}(1984)}]{girvin84}%
  \BibitemOpen
  \bibfield  {author} {\bibinfo {author} {\bibfnamefont {S.~M.}\ \bibnamefont {Girvin}},\ }\bibfield  {title} {\bibinfo {title} {{Anomalous quantum Hall effect and two-dimensional classical plasmas: Analytic approximations for correlation functions and ground-state energies}},\ }\href {https://doi.org/10.1103/PhysRevB.30.558} {\bibfield  {journal} {\bibinfo  {journal} {Phys. Rev. B}\ }\textbf {\bibinfo {volume} {30}},\ \bibinfo {pages} {558} (\bibinfo {year} {1984})}\BibitemShut {NoStop}%
\end{thebibliography}%

\appendix

\setcounter{equation}{0}\renewcommand{\theequation}{A\arabic{equation}}

\section{END MATTER}
\begin{figure}[b!]
    { 
\includegraphics{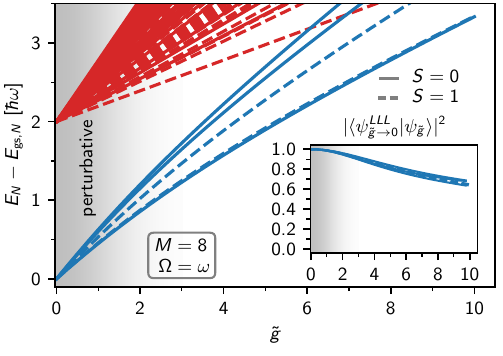}}
    \caption{
    Excitation spectrum as a function of the dimensionless interaction strength $\tilde{g}$ in a rapidly rotating frame \mbox{$\Omega = \omega$} beyond leading-order perturbation theory for \mbox{$N=6$} fermions with angular momentum \mbox{$M=8$}. The shaded region indicates the perturbative regime in which the correspondence between LLL states and states in a harmonic trap holds.
    }    
    \label{fig:4}
\end{figure}

\textit{Stability of LLL states}---\textit{Mixing between LL states at stronger interactions}: The correspondence between LLL states and states in a harmonic trap relies on weak interactions 
\mbox{$\tilde{g}$}. To corroborate the robustness of this correspondence for the states discussed in the main text and to determine the limit of the perturbative region, we calculate the mixing of the LLL states in Fig.~\ref{fig:2}(n) in a restricted exact diagonalization that includes states in the first LL [red states in Fig.~\ref{fig:2}(c)]. Figure~\ref{fig:4} shows the excitation spectrum as a function of the dimensionless interaction strength~\mbox{$\tilde{g}$} for \mbox{$N=6$} fermions with angular momentum \mbox{$M=8$} [we omit the LLL state at \mbox{$M=6$} since it is degenerate with the c.o.m.~excitation]. We show states that evolve from perturbative LLL states in blue and states with an admixture in the first LL in red. The perturbative region is marked by a linear dependence of energies on the coupling constant and a gap to higher excited states of order \mbox{${\it O}(\hbar\omega)$}, which we find extends up to \mbox{$\tilde{g} \simeq 3$} (indicated by the shaded region). 
This is further corroborated in the inset, which shows the overlap of the perturbative LLL states at weak interactions with the lowest-energy eigenstates as a function of $\tilde{g}$. We find that the overlap with the perturbative LLL states remains close to unity, with only a small quadratic correction; for example, for \mbox{$\tilde{g} = 3$} the overlap is still over $90\%$. 
This demonstrates that LLL states are accessible over a broad perturbative region and are robust against stronger interactions.  
\begin{figure}[b!]
    { 
\includegraphics{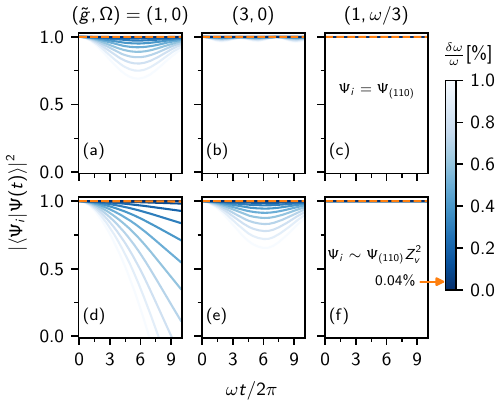}}
    \caption{
    Population of the \mbox{$N=6$} Halperin (1,1,0) state [\mbox{(a)--(c)}] and its (almost) c.o.m.~vortex excitation [\mbox{(d)--(f)}] to leading order due to a static trap anisotropy. The first and second columns are in a nonrotating trap \mbox{$\Omega=0$} with interaction strength \mbox{$\tilde{g}=1$} and \mbox{$\tilde{g}=3$}, respectively, and the third column is in a slowly rotating trap \mbox{$\Omega=\omega/3$} with \mbox{$\tilde{g}=1$}. Results for a state-of-the-art trap anisotropy of \mbox{$\delta \omega /\omega = 0.04\%$}~\cite{lunt24} are shown in orange, for which the states are stable.
    }    
    \label{fig:5}
\end{figure}

\textit{Lifetime of LLL states in a weakly anisotropic trap}: In a perfect harmonic trap, once transferred, the excited LLL states have an infinite lifetime. The main trap imperfection that could induce a finite lifetime and cause a decay to states with lower angular momenta is anisotropy, parametrized by \mbox{$\delta \omega/\omega = (\omega_x-\omega_y)/(\omega_x+\omega_y)$},
with a corresponding perturbation \mbox{$\delta H_{\omega} = m^* \omega \, \delta \omega (x^2-y^2)$}. Such a perturbation couples states that differ by two units in angular momentum. To estimate the stability of the LLL states in the presence of an anisotropy, we show in Fig.~\ref{fig:5} the time evolution of the occupation for two representative states discussed in the main text, the Halperin (1,1,0) state at \mbox{$M=6$} [top row (a)--(c)] and its c.o.m.~vortex excitation at \mbox{$M=8$}  [bottom row (d)--(f)]. Results are obtained using leading-order time-dependent perturbation theory, for which the probability of remaining in an initial eigenstate $|\Psi_i \rangle$ with energy $E_i$ is \mbox{$| \langle\Psi_i|\Psi(t)\rangle|^2 = 1 - \sum_{n\neq i} |c_n(t)|^2$}. Here, $c_n(t)$ are the expansion coefficients in the interaction picture of $|\Psi(t)\rangle $ in the unperturbed Hamiltonian basis~\cite{sakurai94},
\begin{align} \label{eq:amplitude}
c_{n}(t) = \frac{\langle n | \delta H_{\omega} |\Psi_i\rangle }{E_n-E_i} \Bigl( 1 - e^{i(E_n-E_i)t/\hbar} \Bigr),
\end{align}
where $|n\rangle$ and $E_n$ are 
perturbative eigenstates and their energies [Eq.~\eqref{eq:hamiltonian} in the main text]. The first column in Fig.~\ref{fig:5} shows the population for a perturbative interaction strength \mbox{$\tilde{g}=1$} as in Fig.~\ref{fig:2}(a) for trap anisotropies ranging from \mbox{$\delta \omega/\omega = 0$}--$1\%$ (dark to light blue lines). Our time-dependent perturbative analysis~\eqref{eq:amplitude} applies for small-amplitude oscillations, corresponding to small trap anisotropies, while large-amplitude oscillations leave its regime of applicability. In particular, we find that for a state-of-the-art effective trap anisotropy of \mbox{$\delta \omega /\omega = 0.04\%$} [dashed orange line] as determined in Ref.~\cite{lunt24}, all states discussed in the main text are stable. The stability is preserved even when increasing the anisotropy to several times this value. Moreover, even for large anisotropies, the stability can be enhanced considerably by increasing the perturbative level splitting (while not mixing Landau levels), which is demonstrated in the middle column for \mbox{$\tilde{g}=3$} [panels (b) and (e)]. Other ways of rearranging the spectrum at different angular momenta increase the lifetime even further, as illustrated for a slowly rotating trap in the right column [panels (c) and (f), corresponding to Fig.~\ref{fig:2}(b)]. We expect a similar effect for small trap anharmonicity.

\begin{figure}[b!]
    { 
{\includegraphics{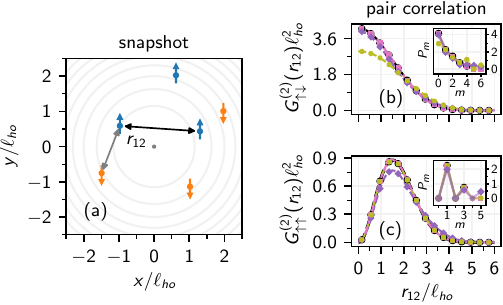}}}
    \caption{(a) Same snapshot of spin-resolved particle positions as in Fig.~\ref{fig:3}(a), this time highlighting interparticle distances $r_{12}$, as indicated here for an opposite-spin (equal-spin) pair [gray (black)]. (b) [(c)] Resulting opposite-spin [equal-spin] pair correlation functions for the four states from Fig.~\ref{fig:3}(b). The insets show the number of pairs $P_m$ in a relative angular momentum channel $m$ as extracted from the pair correlation functions.
    }    
    \label{fig:6}
\end{figure}
\textit{Pair correlation function}---In addition to the observables discussed in the main text, we demonstrate here how the spin-resolved pair correlation function $G^{(2)}_{\sigma \sigma'}(r_{12})$ can be extracted by sampling interparticle distances $r_{12}$, as illustrated in Fig.~\ref{fig:6}(a) by the gray (black) arrow for an opposite-spin (equal-spin) pair. Each such snapshot contains \mbox{$N_{\uparrow}N_{\downarrow}$} opposite-spin and \mbox{$N_{\sigma}(N_{\sigma}-1)/2$} equal-spin pairs, which can be Monte Carlo sampled to obtain $G^{(2)}_{\sigma \sigma'}(r_{12})$. Results for the four states highlighted in the main text are shown as points~\mbox{$G_{\uparrow\downarrow}^{(2)}$} 
[panel~(b)] and \mbox{$G_{\uparrow\uparrow}^{(2)}$} [panel~(c)]. For comparison, we show exact-diagonalization results as dashed curves. The two agree well, which indicates that the sampling of few-particle snapshots provides a reliable means of extracting pair correlation functions. The pair correlation function can be used to extract the number of pairs in a relative angular momentum channel $m$, \mbox{$P_m= \sum_{M'} \langle  \xi_{m M'}^{\dagger} \xi_{m M'}\rangle $}, where $\xi^{\dagger}_{m M'}$ creates a fermion pair in the LLL with relative and c.o.m.~angular momenta $m$ and $M'$~\cite{duncan87}, respectively.
The results are shown in the inset figures, where markers are fits to the Monte Carlo histograms using the expansion \mbox{$\ell_{ho}^2G_{\sigma\sigma'}^{(2)}(r_{12})=\sum_m \frac{P_m}{m!}(r_{12}^2/2\ell_{ho}^2)^{m} e^{-r_{12}^2/2\ell_{ho}^2}$}~\cite{girvin84}, and the lines are computed from exact diagonalization. Only odd $m$ enter the expansion of $G_{\sigma\sigma}^{(2)}$ for fermions, and the interaction energy in units of $\hbar\omega$ is given by $\frac{\tilde{g}}{2\pi} P_0$ for contact interactions in the LLL. While the pair correlation functions take a similar shape for the states near \mbox{$\nu=2$} discussed here, they will be sensitive to interaction effects for states at smaller filling fraction (corresponding to larger $M$).

\end{document}